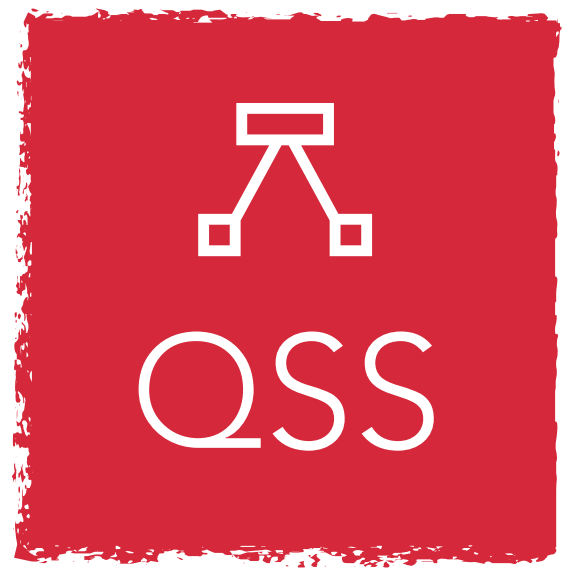

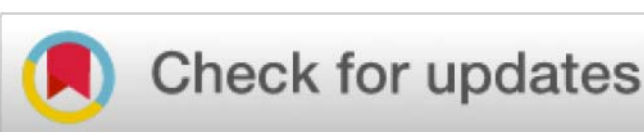

Corresponding Author:
Marek Kwiek
kwiekm@amu.edu.pl

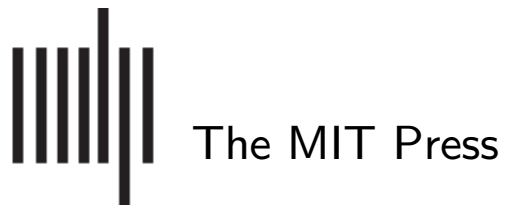

# Quantifying lifetime productivity changes: A longitudinal study of 320,000 late-career scientists

**Marek Kwiek**[1] and **Lukasz Szymula**[2,3]

[1]Center for Public Policy Studies, Adam Mickiewicz University, Poznan, Poland
[2]Department of Artificial Intelligence, Adam Mickiewicz University, Poznan, Poland
[3]Department of Computer Science, University of Colorado Boulder, Boulder, CO, USA

## ABSTRACT

The present study focuses on persistence in research productivity over the course of an individual's entire scientific career. We track "late-career" scientists—scientists with at least 25 years of publishing experience ($N$ = 320,564)—in 16 STEMM (science, technology, engineering, mathematics, and medicine) and social science disciplines from 38 OECD countries for up to 5 decades. Our OECD sample includes 79.42% of late-career scientists globally. We examine the details of their mobility patterns as early-career, midcareer, and late-career scientists between decile-based productivity classes, from the bottom 10% to the top 10% of the productivity distribution. Methodologically, we turn a large-scale bibliometric data set (Scopus raw data) into a comprehensive, longitudinal data source for research on careers in science. The global science system is highly immobile: Half of global top performers continue their careers as top performers and one-third of global bottom performers as bottom performers. Jumpers-Up and Droppers-Down are extremely rare in science. The chances of moving radically up or down in productivity classes are marginal (1% or less). Our regression analyses show that productivity classes are highly path-dependent: There is a single most important predictor of being a top performer, which is being a top performer at an earlier career stage.

## 1. INTRODUCTION

The focus of the present study is persistence in top and bottom individual research productivity from a lifetime perspective—that is, over the course of an entire scientific career. We are tracking "late-career" scientists ($N$ = 320,564) from 38 OECD countries for up to 5 decades to examine their mobility patterns between decile-based productivity classes, from the bottom 10% to the top 10%.

We turn a large-scale publication and citation bibliometric data set (Scopus raw data) into a global, comprehensive, multidimensional, and longitudinal data source for research on careers in science; this practice follows several previous global studies on gender and self-citations (King, Bergstrom et al., 2017), women in science (Huang, Gates et al., 2020; Larivière, Ni et al., 2013; Sugimoto & Larivière, 2023; West, Jacquet et al., 2013), global citation inequality (Nielsen & Andersen, 2021), continuing publishing core of global science (Ioannidis, Boyack, & Klavans, 2014), collaboration with top scientists (Li, Aste et al., 2019), the nature of international collaboration (Wagner, 2018), or academic careers viewed from the perspective of science of science (Wang & Barabási, 2021). So far, however, individual publishing productivity has not been examined globally from a longitudinal perspective.

Because our study is of a longitudinal nature, we use a global bibliometric data set to study scientific careers and how they change over time (Menard, 2002; Rowland, 2014; Ruspini, 1999): The same individuals are tracked over the multiple decades of their publishing careers. In global academic career research, ever more data sets are currently tested (e.g., integrated data sets with administrative and biographical, commercial and noncommercial, national and global data; see, for example, King et al., 2017; Larivière et al., 2013; Nielsen & Andersen, 2021). In the present research, we test the usefulness of publication and citation metadata for examining the global science profession from a longitudinal perspective; these metadata are digital traces left by scientists throughout their professional lives (or as long as they keep publishing in academic journals). Digital traces (Liu, Jones et al., 2023; Salganik, 2018) allow for the emergence of a whole new multidisciplinary field of science of science (Clauset, Larremore, & Sinatra, 2017; Wang & Barabási, 2021; Zeng, Shen et al., 2017), hence allowing science career studies to radically move beyond traditional small-scale surveys and interviews (Hermanowicz, 2012; Leišytė & Dee, 2012). The digital traces left by scientists in global publication and citation data sets allow academic career researchers to change their focus from single national science systems to a global science system (Huang et al., 2020; King et al., 2017; Ni, Smith et al., 2021).

Our initial hypotheses, which are based on research productivity literature (Allison, Long, & Krauze, 1982; Fox, 1983; Turner & Mairesse, 2005), especially high research productivity literature focused on "top performers" and "prolific" scientists (Abramo, D'Angelo, & Caprasecca, 2009; Aguinis & O'Boyle, 2014; Fox & Nikivincze, 2021; Kwiek, 2016; Li et al., 2019) are, first, that scientists are generally locked in within their productivity classes for years (Kelchtermans & Veugelers, 2013; Turner & Mairesse, 2005); second, we argue that the elite strata of highly productive scientists often continue their entire careers as being highly productive (Allison & Stewart, 1974; David, 1994); and, finally, we argue that radical changes in productivity classes, especially upward (Jumpers-Up), although present in popular narratives about academic careers, are highly improbable in practice because of the cumulative nature of the advantages and disadvantages in careers, as shown over the decades in the traditional sociology of science (Cole & Cole, 1973; DiPrete & Eirich, 2006; Merton, 1973).

The current study follows the research lines explored in the field of science of science, which provides data-driven insights into the inner workings of science (Wang & Barabási, 2021). A shift toward new digitalized data sources allows for the exploration of new questions about scientists (Liu et al., 2023). In this case, traditional cross-sectional studies can be complemented with longitudinal studies (Lutter & Schröder, 2016; Ma, Mukherjee, & Uzzi, 2020) in which individuals are tracked over time.

The career histories of thousands of individual scientists can change the way we think about science and scientists because of the unprecedented level of detail that can be obtained. As a result, the various aspects of academic careers have recently been examined both globally (gender disparities in careers: Huang et al., 2020; Larivière et al., 2013; continuous publishing: Ioannidis et al., 2014; collaboration with top scientists: Li et al., 2019; gendered nature of authorship: Ni et al., 2021; women in science: Sugimoto & Larivière, 2023) and nationally, especially in the United States (e.g., productivity across career stages: Way, Morgan et al., 2017; long-term effects on careers of initial setbacks: Wang, Jones, & Wang, 2019; careers in elite universities: Zhang, Wapman et al., 2022; credit distribution in academic publishing: Ross, Glennon et al., 2022) at a scale unthinkable in career studies before.

### 1.1. Productivity Classes in Single-Nation Studies

We have tested our initial hypotheses in a previous national-level strand of research under the general label of "once highly productive, forever highly productive" (Kwiek & Roszka, 2024a, 2024b). The patterns found about OECD scientists consistently supported our initial intuitions about immobility in the system based on our single-nation research: The majority of highly productive early-career scientists and midcareer scientists continued their careers as highly productive midcareer scientists and late-career scientists; radical upward and downward mobility was either at zero or at marginal levels. In the present research, we develop our methodological approach to study mobility between the productivity classes of scientists from 38 OECD countries that are often powerfully involved in international collaboration as part of the ongoing globalization of science (Kwiek, 2023; Marginson, 2022). We track 320,564 late-career scientists from a wide variety of research systems, giving us the potential to test hypotheses about the scientific profession more generally.

There are three small-scale longitudinal single-nation studies similar to ours. First, for 497 French physicists, Turner and Mairesse (2005) showed that 66% of the most productive researchers (defined as quartile 1 scientists) and 67% of the least productive researchers (defined as quartile 4 scientists) remained as such for the period 1986–1997, underlying the stability of the relative positions of the researchers in the distribution of publication counts over time. Second, in a study of a single Belgian university, Kelchtermans and Veugelers (2013) discussed top research productivity and its persistence over time by using a panel data set comprising the publications of 1,040 biomedical and exact scientists for the period 1992–2001; they studied how researchers switch between productivity categories over time, showing strong support for an accumulative process. However, this places scientists with a low initial output at a disadvantage, while giving highly productive scientists a greater advantage. Abramo, D'Angelo, and Soldatenkova (2017) studied Italian scientists in three consecutive 4-year periods of 2001–2012; they identified 2,883 top performers in the first period and followed them over time. About one-third of top performers retained their top ranking for three consecutive periods, and about half retained it for two periods (35% and 55%, respectively).

Our research explores a different scale, scope, and methodology: We track a large number of late-career scientists from 38 OECD countries from all science sectors (including higher education); we examine productivity changes over a prolonged period (25–50 years) across all STEMM and selected social science disciplines; and we use a longitudinal and classificatory approach combined with two-dimensional analyses and logistic regression models.

### 1.2. Persistence in High (and Low) Productivity

Productive scientists are likely to be "even more productive in the future, while scientists who produce little original work are likely to decline further in their productivity" (Allison & Stewart, 1974, p. 596). Substantial predetermined differences among scientists may have a powerful impact on careers (Cole & Cole, 1973; Fox, 1983). An "initial success" may lead to increased productivity; in contrast, a "bad start" may lead to the scientist leaving science (Turner & Mairesse, 2005). In other words, "there are substantial, predetermined differences among scientists in their ability and motivation to do creative scientific research" (Allison & Stewart, 1974, p. 596). Some scientists are always very productive, and a differential distribution of talent affects inequality in productivity more than the recognition system in science (Stephan & Levin, 1992).

As a result, productivity stratification in science leads to "persistent hierarchies of productivity": "once scientists enter the current productivity elite, it is rare for them to exit from it in the next period; and the same holds true at the lower extreme of the productivity distribution" (David, 1994). Top performers tend to try hard not to disappoint their colleagues and themselves; bottom performers, in contrast, tend to lose confidence in their research capabilities. Previous top performance significantly and positively affects current top performance (Kelchtermans & Veugelers, 2013).

### 1.3. Research Questions

At a global scale, we quantify the persistence in research productivity over an academic lifetime by following previous research (using analyses based on small-scale surveys and a limited number of interviews) across a span of decades (Allison, 1980; Allison et al., 1982; Cole & Cole, 1973; Merton, 1973). Tracking the career trajectories of thousands of scientists, we seek global mobility patterns between research productivity classes (whenever we use the term "global," we refer to 38 OECD countries). Our sample of OECD late-career scientists includes 79.42% of all late-career scientists globally (from all countries, $N$ = 403,653); and their research output (30,695,679 research articles) includes 83.03% of all research articles produced by this category of scientists globally ($N$ = 36,969,473).

Consequently, we have posed the following research questions regarding changing publishing productivity over the course of individuals' academic careers: First, what is the scale of horizontal transitions (top to top, bottom to bottom) and radical vertical transitions (bottom to top, top to bottom) between decile-based global productivity classes? Second, what is the scale of jumping up (and dropping down) in science in terms of research productivity—radically changing productivity classes upward or downward globally? Third, what are the cross-disciplinary differences in mobility patterns between global productivity classes? Finally, what are the predictors of belonging to the classes of highly productive and bottom productive scientists (the top 10%, the bottom 10%), and how do they differ between academic disciplines?

## 2. METHODOLOGY

In this research, we move from individual publications (and their properties) to individual scientists (and their characteristics) as a unit of analysis. We construct individual lifetime publication and citation histories for every late-career scientist in our sample, restricting our research to 16 STEMM (science, technology, engineering, mathematics, and medicine) and social science disciplines. In our context, "late-career" scientists are defined as scientists with at least 25 years of publishing experience.

We explore mobility between the 10 individual productivity classes (constructed according to the 10 decile-based classes) throughout long academic careers, here encompassing early-, mid-, and late-career periods.

### 2.1. Data

The data were collected from the Scopus bibliometric database and were obtained through a multiyear collaborative agreement with the International Center for the Study of Research (ICSR) Lab, a cloud computing platform provided for research purposes by Elsevier. Our final sample included all late-career scientists who were research active in 2023 (with at least 25 years of publishing experience) located in 16 STEMM and social science disciplines and coming from 38 OECD countries ($N$ = 320,564 scientists with $N$ = 16,345,891 research articles;

see the major steps in data preprocessing in Figure 1; the "authors" in our data set were defined as having publications of any type and scientists as authors with articles in journals and conference proceedings only. For our calculations, we utilized the Scopus database dated March 29, 2024.

Our sample (Tables 1 and 2 in the Electronic Supplementary Material [ESM]) included 12,585 social scientists (from BUS, ECON, and PSYCH) and 307,979 STEMM scientists, the latter comprising 95.76% of our sample. About a quarter of our sample included women scientists (26.34%), the number of which was slightly more in the social sciences than STEMM fields. The largest academic discipline represented in our sample was MED (40.89%), followed by BIO (14.29%) and PHYS (9.13%). The percentage of women in our sample was about one-third in three disciplines (the two largest: MED and BIO) and exceeded 40% in only one (PSYCH: 41.44%). The three largest countries represented were the United States, Japan, and Italy, comprising about a half of all scientists in the sample.

In terms of academic age, there are about 20,000–25,000 scientists in the youngest cohorts and about 2,000–3,000 in the oldest cohorts, with the share of women decreasing with each subsequent cohort, from about one-third for the youngest cohort (25 years of academic experience: 32.77%) to 13–15% for the oldest cohorts. In practical terms, we are working with the census of a population with clearly defined inclusion criteria rather than with a sample of scientists—with methodological implications for testing for statistical significance (which are not needed in the present research). The probability that the observed relationships and differences occurred by pure chance are zero because we do not work with samples drawn from the population of late-career scientists but with their full population.

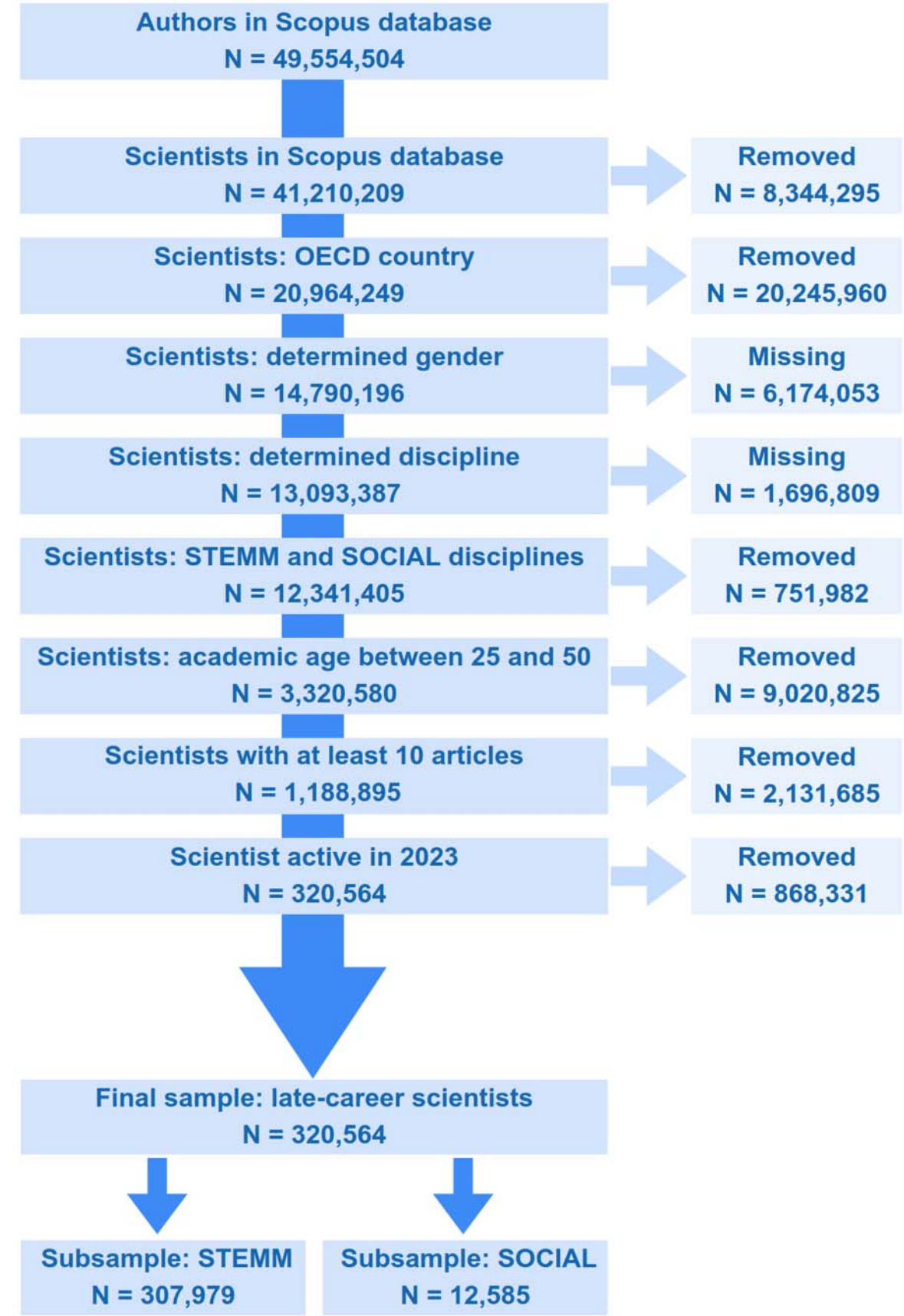

**Figure 1.** Flowchart and major steps in data preprocessing: from all scientists in the Scopus database to late-career scientists in our sample.

To achieve aggregate-level results, the ICSR Lab employed the Databricks environment, which facilitates the management and execution of cloud computing with Amazon EC2 services. The scripts for generating the results were developed using the PySparkSQL library. Runs were carried out using a cluster in standard mode with Databricks Runtime version 11.2 ML, Apache Spark technology version 3.3.0, Scala 2.12, and an i3.2xlarge instance with 61 GByte memory, eight cores, one to six workers for the worker type, and a c4.2xlarge instance with 15 GByte memory and four cores for the driver type. The execution time was six hours, and this operation was initiated on June 25, 2024. We obtained the results in CSV format.

### 2.2. Methods

The academic lives of all late-career scientists from 38 OECD countries that were research active in 2023 were retrospectively divided into three stages: early-, mid-, and late-career stages. All late-career scientists, by definition, were initially both early-career scientists (in their publishing years 5–14) and midcareer scientists (in their publishing years 15–24). We analyzed their current 5-year publishing behavior (2019–2023) and looked back into their past publishing behavior to examine how they may have changed their productivity classes.

At each career stage, current late-career scientists showed their annual individual productivity. Consequently, their productivity was calculated for the recent 5-year period and for two earlier periods: when they were early-career scientists and midcareer scientists. Our analyses are based on the idea of subsequent distributions of scientists into classes: Late-career scientists are first distributed by current productivity classes (separately within each of the 16 disciplines) and then, retrospectively, by past productivity classes in the two earlier career periods.

Early-career scientists may retain or change their decile-based productivity classes while being midcareer scientists, as midcareer scientists may do while being late-career scientists. In the present study, we tracked scientists for 25–50 years and compared their productivity with the productivity of their peers (the same academic career stage and the same discipline).

For each scientist in our sample, an individual publication and citation portfolio was constructed. The portfolio included Scopus-derived publication metadata and their various constructs that accompanied individual authors from their first publication in the data set to the year 2023. Within the portfolios, all metadata and their specially computed constructs were linked to the three career periods (e.g., annual productivity), individual publications (e.g., field-weighted 4-year citation impact), or the entire lifetime careers of scientists (e.g., gender, discipline, international collaboration rate, and median team size) (see Variables in Table 1).

Our approach to individual research productivity is longitudinal (Menard, 2002; Rowland, 2014; Ruspini, 1999) and classificatory (or class based) (Costas & Bordons, 2007; Costas, van Leeuwen, & Bordons, 2010). First, we tracked the productivity of late-career scientists as individuals ever since they became early-career scientists, that is, 5 years after their first globally indexed publication. Second, we did not compare productivity changes over time (as individual scientific careers develop) in terms of changing publication numbers—we compared productivity in terms of the stable or changing membership in the productivity classes while scientists grow older and move up the professional ladder. Scientists can always be allocated to the top and bottom classes, regardless of actual publication numbers, cut-off points permitting, so that both terms could be used not to judge the level of productivity but rather to classify it.

**Table 1.** Variables used in regression analysis

| No. | Variable | Description |
|---|---|---|
| 1. | Gender | Gender (binary: female/male) provided by Elsevier's ICSR Lab. Variable classified based on the first name, last name, and dominant country from the first year of publishing using the Namsor tool. Gender accepted with the probability score ≥ 0.85 only. |
| 2. | Field-weighted 4-year citation impact (FWCI 4y) | Average of the FWCI 4y metric values assigned to each publication in author's lifetime publication portfolio. The FWCI 4y metric value of a publication means the ratio of the number of citations of that publication (obtained in the publication year and 3 consecutive years) to the average number of citations for a similar publication (publication from the same discipline group in the Scopus four-digit ASJC discipline classification) in the same time frame. |
| 3. | International collaboration rate (lifetime) | Share of author's international collaborative publications among all collaborative publications (solo publications excluded). For a publication to be considered collaborative, the number of all authors in the paper had to be greater than or equal to two. For a publication to be considered international, the number of affiliation countries in the paper had to be greater than or equal to two. |
| 4. | Median team size (lifetime) | Median of the number of authors for each publication (author + number of collaborators) in author's lifetime publication portfolio. For publications with the number of authors greater than 10, the number of authors is 10. |
| 5. | Discipline | Dominant discipline based on the modal value from all disciplines assigned to the journals of all cited references in all papers in scientists' lifetime publication portfolios. |
| 6. | TOP200 institutional affiliation | Binary value indicating belonging (true/false) to one of the 200 top institutions. The list of top institutions was ranked based on the institutions' total scholarly output in the 10-year period 2014–2023. Each author has been assigned to one institution as the dominant one based on the modal value from all institutions indicated in an author's lifetime publication portfolio. Variable used only for second transitions: midcareer to late career (affiliation in early career is too distant in time). |
| 7. 8. | Early career/midcareer top class | Membership in the global top 10% of scientists among early-career/midcareer scientists in terms of research productivity (journal prestige-normalized, full counting, exponential function), separately within 16 disciplines. |
| 9. 10. | Early career/midcareer bottom class | Membership in the global bottom 10% of scientists among early-career/midcareer scientists in terms of research productivity (journal prestige-normalized, full counting, exponential function), within 16 disciplines. |

We used a journal prestige-normalized, full counting method of calculating productivity (see ESM for more details). This approach refers to the quantity and quality of globally indexed publications at the level of individuals (as opposed to quantity only in nonnormalized approaches). Prestige normalization refers to the journal percentile ranks used in the Scopus database (CiteScore ranking, range: 1–99), and it highlights the difference in average scholarly efforts between preparing and revising publications in generally less selective and more selective journals, with different peer review procedures and acceptance rates. Prestige normalization is determined by the number of citations received by the journal (43,092 journals in 2024) in the previous 4 years. In a prestige-normalized approach, the weight of publications depends on their location in a vertically stratified system of academic journals (for more on the role of journal stratification in academic careers, see Hammarfelt, 2017; Heckman & Moktan, 2018; Kwiek, 2021; Lindahl, 2018; Shibayama & Baba, 2015).

### 2.3. Variables and Their Measures

Importantly, our approach does not rely on publication numbers because productivity across OECD countries has generally increased over the past decades, especially when computed on a full counting rather than fractional counting basis. Instead, we rank all current late-career scientists by productivity, assigning them to specific productivity classes within their respective academic disciplines. We then retrospectively rank these late-career scientists based on their productivity during their early- and midcareer stages by using 4-year periods to measure their productivity at these times.

Our focus is on analyzing the mobility between productivity classes, particularly the transitions between the top and bottom classes and the adjacent classes: productivity deciles 8, 9, and 10 at the top and deciles 1, 2, and 3 at the bottom of the productivity distribution. We examine three stages of academic careers—early career, midcareer, and late career—and the transitions from early career to midcareer and from midcareer to late career.

Early-career scientists in the top and bottom productivity classes can change their productivity classes as they progress to the midcareer stage, moving to any decile. Similarly, midcareer scientists in the top and bottom productivity classes can experience changes in their productivity levels as they transition to the late-career stage, moving up, down, or staying in the same productivity decile. We want to understand how productivity can evolve over the course of an academic career and discuss the extent to which individuals move between different levels of productivity.

We examine three primary types of mobility (across all academic disciplines and within specific disciplines):

1. *Top-to-top mobility*: Early-career scientists who are in the highest productivity decile remain in the highest decile as they progress to the midcareer stage, and similarly, from the midcareer to late-career stages. This reflects persistence in high productivity from one career stage to the next (mobility from decile 10 to decile 10).
2. *Bottom-to-bottom mobility*: Early-career scientists in the lowest productivity decile stay in the lowest decile as they advance to the midcareer stage and, likewise, from the midcareer to late-career stages. This indicates a persistent low productivity level across career stages (mobility from decile 1 to decile 1).
3. *Extreme downward and extreme upward mobility* (*top-to-bottom mobility* and *bottom-to-top mobility*): This includes both downward and upward mobility. Scientists who start in the top productivity decile in their early careers and who drop to the bottom decile by the mid- or late-career stage (top-to-bottom mobility, Droppers-Down); and scientists who begin in the lowest decile and rise to the highest by mid- or late-career stage (bottom-to-top mobility, Jumpers-Up). This represents significant shifts in productivity, either from decile 10 to decile 1 or from decile 1 to decile 10.

In addition to examining the basic mobility between the highest (decile 10) and lowest (decile 1) productivity deciles, we will also explore a broader perspective of mobility that considers the transition between the upper deciles (8–10) and lower deciles (1–3). Some scientists are positioned just above the decile 1 threshold and others just below the decile 10 threshold (see Tables 3, 4, and 5 in ESM). A more comprehensive approach that includes adjacent deciles (1–3 and 8–10) seems useful.

Our general question is how top-performing (productivity decile 10, $N$ = 32,063) midcareer scientists were distributed by productivity percentile ranks (range: 0–100) when they were early-career scientists in the past and, analogously, how top-performing (productivity decile 10, $N$ = 32,063) late-career scientists were distributed by productivity percentile ranks when they were midcareer scientists.

In addition, we are interested in how bottom-performing (productivity decile 1, $N$ = 32,063) midcareer scientists were distributed by productivity percentile ranks (range: 0–100) when they were early-career scientists in the past. In addition, analogously, how the current bottom-performing (productivity decile 1, $N$ = 32,075) late-career scientists were distributed by productivity percentile ranks when they were midcareer scientists. In all cases studied, we retrospectively examine current late-career scientists (who make up our sample): when they were early-career scientists and when they were midcareer scientists.

In the second part of Section 3, we use regression models to estimate the odds ratios of membership in the classes of global top productive and bottom productive midcareer and late-career scientists (the top 10%, separately for each academic discipline). The variables used in regression analysis are described in Table 1.

In the present research, we have used a global bibliometric data set to define scientists' individual attributes. The determination of some attributes has already been described in detail in our previous research: gender determination (binary: male or female), discipline determination (using all cited references from all publications, lifetime), determining the country of affiliation (using a modal value of all affiliations in all publications, lifetime), determination of scientists' nonoccasional status in global science (using a minimum output of 10 research articles), and determining academic age (using the distance in years between the first publication, of any type, and 2023; Kwiek & Szymula, 2023, 2024). Three other individual attributes were used in individual publication and citation portfolios (their construction is described in Table 1): international collaboration rate (lifetime), field-weighted 4-year citation impact (FWCI 4y), and median team size (lifetime). The distribution of the sample by academic age (i.e., publishing experience) is shown in Figure 2, with further details in Tables 1 and 2 in ESM.

### 2.4. Means of Analysis and Presentation

In this research, for presentations, we use kernel density plots to show the distributions of top-performing and bottom-performing scientists by productivity percentiles, in the range 0–100); and we use Sankey diagrams to show scientists' mobility between productivity classes in the three stages of a scientific career. Kernel density plots utilize kernel density estimation to generate a smooth, continuous curve that represents the underlying data distribution. Unlike histograms, kernel density plots are not influenced by the number of bins or significant differences between them, making them more effective in illustrating the shape of a distribution; they also allow for a more flexible comparison between multiple data sets. Sankey diagrams are used to compare amounts through different stages and they are a data visualization technique that emphasizes flow from one state to another—in which the width of the flows is proportional to the flow rate. Thicker flows indicate larger proportions than thinner flows.

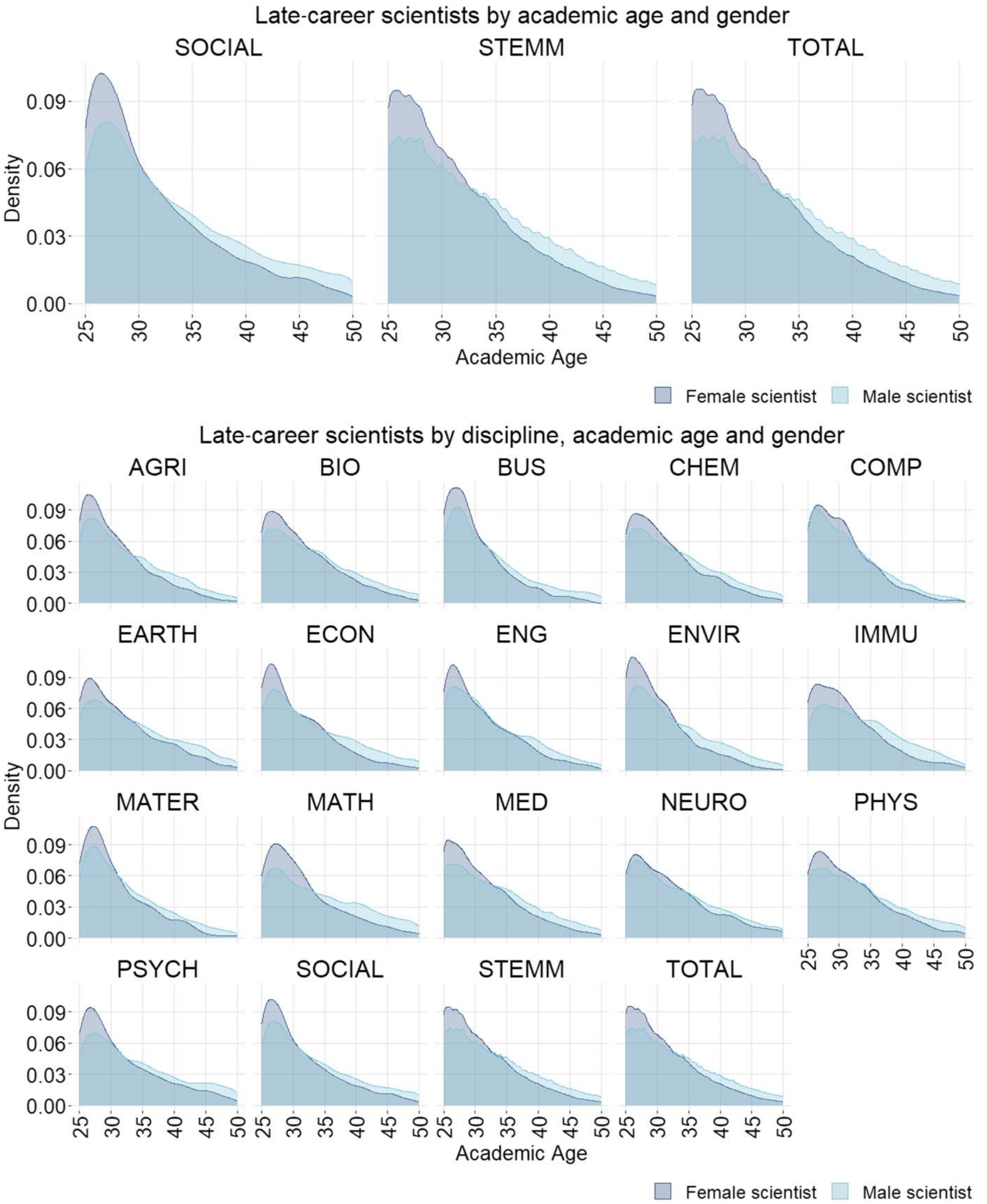

**Figure 2.** Distribution of academic age: Kernel density plots. Late-career scientists, all academic disciplines combined (top panel) by gender. Late-career scientists by academic discipline (bottom panel) and gender ($N$ = 320,564).

## 3. RESULTS

### 3.1. Changing Productivity Percentiles Ranks Over Time

In the mobility from the early-career to midcareer stage, the median value of the original percentile rank (as early-career scientist) is very close to the target percentile rank (as midcareer scientist): The median is the 90th percentile for global top performers and the 15th percentile for global bottom performers (Table 2), with limited discipline-related variability for top performers (from 89th in COMP and ENG to 92nd in MATH) and slightly higher discipline-related variability for bottom performers (from 12th in PHYS to 17th in COMP and ECON). Differences in both the median and mean values between social science academic disciplines combined (SOCIAL) and STEMM disciplines combined are limited.

**Table 2.** How top performers at a midcareer stage (productivity decile 10) (Left panel) and bottom performers at a midcareer stage (productivity decile 1) (Right panel) were distributed by productivity percentiles (range: 0–100) when they were in their early-career stage. Top/bottom performers in their midcareer stage, initial (as early-career stage) percentile distribution statistics by academic discipline ($N_{top}$ = 32,063, $N_{bottom}$ = 32,063)

| | (Left panel) Top performers at a midcareer stage—distribution at an early-career stage | | | | (Right panel) Bottom performers at a midcareer stage—distribution at an early-career stage | | | |
|---|---|---|---|---|---|---|---|---|
| **Discipline** | ***N*** | **Mean** | **Std dev** | **Median** | ***N*** | **Mean** | **Std dev** | **Median** |
| AGRI | 2,373 | 85.03 | 15.84 | **91** | 2,373 | 18.66 | 17.13 | **14** |
| BIO | 4,582 | 82.95 | 18.69 | **90** | 4,582 | 21.05 | 19.58 | **15** |
| BUS | 326 | 82.00 | 19.74 | **89.5** | 326 | 24.43 | 21.44 | **18** |
| CHEM | 1,490 | 84.52 | 17.55 | **91** | 1,490 | 17.57 | 16.68 | **13** |
| COMP | 765 | 80.80 | 20.73 | **89** | 765 | 23.12 | 20.07 | **17** |
| EARTH | 1,437 | 85.01 | 16.27 | **91** | 1,437 | 18.32 | 17.48 | **13** |
| ECON | 385 | 82.32 | 19.85 | **90** | 385 | 22.30 | 19.20 | **17** |
| ENG | 1,282 | 82.40 | 18.84 | **89** | 1,282 | 20.85 | 19.02 | **15** |
| ENVIR | 652 | 84.82 | 16.78 | **91** | 652 | 20.99 | 19.48 | **15.5** |
| IMMU | 315 | 82.28 | 19.21 | **89** | 315 | 21.24 | 20.46 | **14** |
| MATER | 584 | 83.26 | 17.91 | **90** | 584 | 17.89 | 16.18 | **14** |
| MATH | 701 | 85.58 | 16.79 | **92** | 701 | 19.90 | 18.01 | **15** |
| MED | 13,108 | 83.97 | 17.04 | **90** | 13,108 | 20.15 | 18.49 | **15** |
| NEURO | 587 | 83.72 | 18.50 | **90** | 587 | 19.90 | 18.88 | **14** |
| PHYS | 2,928 | 83.25 | 20.90 | **92** | 2,928 | 18.49 | 19.15 | **12** |
| PSYCH | 548 | 83.60 | 18.35 | **91** | 548 | 18.79 | 17.74 | **13** |
| **SOCIAL** | **1,259** | **82.79** | **19.17** | **90** | **1,259** | **21.32** | **19.14** | **17** |
| **STEMM** | **30,804** | **83.78** | **17.77** | **90** | **30,804** | **19.88** | **18.53** | **15** |
| **TOTAL** | **32,063** | **83.74** | **17.83** | **90** | **32,063** | **19.94** | **18.55** | **15** |

Similarly, regarding the mobility from the midcareer to late-career stage, the median value of the initial percentile rank (as midcareer scientist) for the current late-career top performers is very close to the target percentile rank (as late-career scientist): The median is the 90th percentile for top performers and the 19th percentile for bottom performers (Table 3), with limited discipline-related variability for top performers and slightly higher discipline-related variability for bottom performers (from 15th in PSYCH to 21st in BIO and ECON). Again, the differences in both the median and mean values between social science academic disciplines (SOCIAL) and STEMM academic disciplines combined are marginal.

A useful way to visualize the distribution of current top-performing and bottom-performing late-career scientists across productivity deciles during their midcareer and early-career stages is presented through kernel density plots (Figures 3 and 4). When considering all academic disciplines combined (TOTAL), the majority of top performers were previously in productivity deciles 8 through 10, while most bottom performers were previously in deciles 1 through 3. Notably, the concentration of top-performing midcareer and late-career scientists is considerably higher in all STEMM disciplines combined than in all Social disciplines combined. The patterns of distribution are stunningly similar across disciplines and career stages. Initial percentile distributions are sharper for top-performing scientists, with MED clearly reflecting the overall pattern for all disciplines examined.

**Table 3.** How top performers in their late-career stage (productivity decile 10) (Left panel) and bottom performers in their late-career stage (productivity decile 1) (Right panel) were distributed by productivity percentiles (range: 0–100) when they were in their midcareer stage. Top/bottom performers in a late-career stage, initial (as midcareer stage) percentile distribution statistics by academic discipline ($N_{top}$ = 32,063, $N_{bottom}$ = 32,075)

| | (Left panel) Top performers at a late-career stage—distribution at a midcareer stage | | | | (Right panel) Bottom performers at a late-career stage—distribution at a midcareer stage | | | |
|---|---|---|---|---|---|---|---|---|
| **Discipline** | ***N*** | **Mean** | **Std dev** | **Median** | ***N*** | **Mean** | **Std dev** | **Median** |
| AGRI | 2,373 | 84.34 | 17.98 | **91** | 2,373 | 21.90 | 19.08 | **17** |
| BIO | 4,582 | 82.97 | 19.52 | **90** | 4,582 | 26.37 | 22.31 | **21** |
| BUS | 326 | 80.33 | 21.99 | **89** | 326 | 25.52 | 20.82 | **20** |
| CHEM | 1,490 | 85.20 | 17.92 | **92** | 1,490 | 21.92 | 19.39 | **16** |
| COMP | 765 | 81.76 | 21.08 | **90** | 767 | 24.55 | 20.68 | **19** |
| EARTH | 1,437 | 82.50 | 19.38 | **90** | 1,437 | 21.70 | 19.70 | **15** |
| ECON | 385 | 80.38 | 22.31 | **90** | 385 | 25.44 | 20.76 | **21** |
| ENG | 1,282 | 83.06 | 19.74 | **91** | 1,282 | 23.38 | 19.26 | **19** |
| ENVIR | 652 | 83.58 | 19.25 | **91** | 652 | 24.69 | 20.37 | **19** |
| IMMU | 315 | 82.78 | 19.81 | **90** | 319 | 24.12 | 22.07 | **17** |
| MATER | 584 | 85.03 | 17.84 | **91** | 590 | 22.17 | 18.93 | **17** |
| MATH | 701 | 84.21 | 19.00 | **91** | 701 | 22.07 | 19.02 | **18** |
| MED | 13,108 | 83.01 | 19.61 | **90** | 13,108 | 24.50 | 21.09 | **19** |
| NEURO | 587 | 84.53 | 18.35 | **91** | 587 | 24.55 | 21.75 | **18** |
| PHYS | 2,928 | 83.28 | 21.94 | **92** | 2,928 | 22.37 | 20.84 | **16** |
| PSYCH | 548 | 83.84 | 18.10 | **91** | 548 | 21.29 | 19.69 | **15** |
| **SOCIAL** | **1,259** | **81.87** | **20.39** | **90** | **1,259** | **23.65** | **20.31** | **20** |
| **STEMM** | **30,804** | **83.29** | **19.56** | **90** | **30,816** | **23.98** | **20.78** | **19** |
| **TOTAL** | **32,063** | **83.23** | **19.60** | **90** | **32,075** | **23.96** | **20.76** | **19** |

### 3.2. Changing Productivity Deciles Over Time: Jumpers-Up and Droppers-Down

With our data set, we can analyze the mobility between productivity deciles (at the level of individuals) in great detail. Table 4 shows initial productivity deciles (as early-career scientists) in the past of top-performing midcareer scientists across the various academic disciplines.

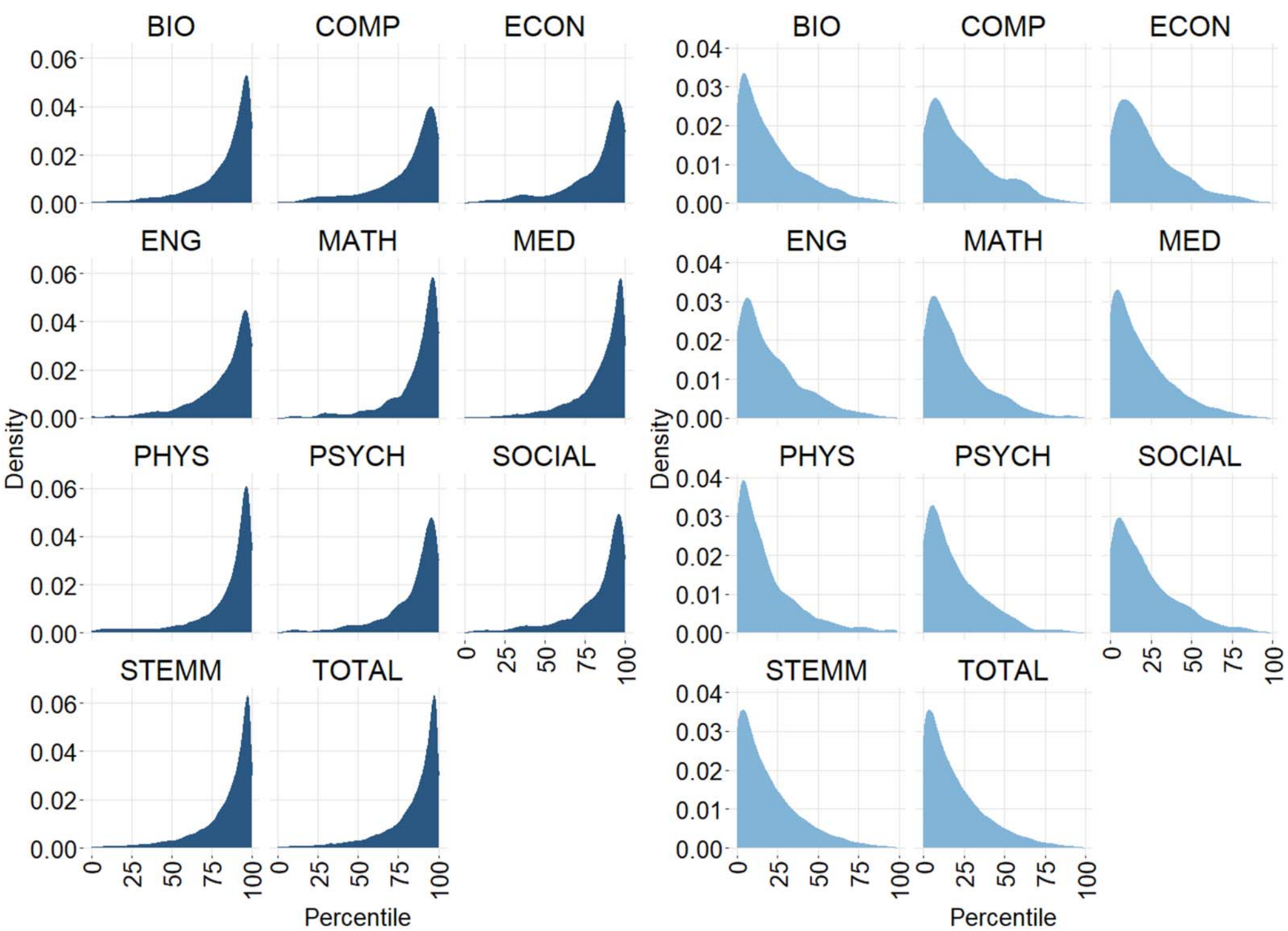

**Figure 3.** (Left panel) How were the top-performing ($N$ = 32,063, productivity decile 10) midcareer scientists distributed by productivity percentiles (range: 0–100) when they were in their early-career stage? (Right panel) How were the bottom-performing ($N$ = 32,063, productivity decile 1) midcareer scientists distributed by productivity percentiles (range: 0–100) when they were in their early-career stage? Kernel density plots, initial percentile distribution, by selected discipline.

Over half of these top midcareer scientists were in productivity decile 10 in their early-career stage (52.39%), with 20.94% starting off in decile 9 and 10.33% decile 8. Altogether, more than 80% were in productivity deciles 8–10 during their early-career stage (83.66%).

Only a small fraction of these scientists moved up from the lowest three deciles, with just 162 making a significant leap from decile 1 to decile 10 (0.51%, referred to as Jumpers-Up) and 232 from decile 2 to decile 10 (0.72%). We have full lifetime biographical and publishing profiles of every scientist, including these few hundreds of outliers. Overall, only 2.2% (717 scientists out of 32,063) from deciles 1–3 reached decile 10. In social science disciplines, the likelihood of such extreme upward mobility was slightly higher compared with STEMM disciplines (3.13% vs. 2.22%) but still relatively rare.

The variation in extreme mobility from decile 1 to decile 10 (Jumpers-Up) across disciplines is significant, with rates ranging from 0.26% in ECON to 1.33% in PHYS. Only one economist (0.26%) and one immunologist (0.32%) made the leap from decile 1 to decile 10 (out of 385 and 315, respectively). We know a high number of details of their specific academic careers based on our bibliometric data (but not their scientific biographies based on administrative data from national registries of scientists as in single-nation studies—not available for a multi-country study). The mobility from decile 10 to decile 10 shows variability as well, with slightly less than 50% of scientists in COMP and ENG remaining in decile 10 compared with 57–58% in MATH and PHYS.

Similarly, when examining the mobility from midcareer to late career for top-performing scientists (Table 5), the persistence in top productivity is even more pronounced, with 53.83% of scientists who started in decile 10 remaining in decile 10. Only one in six (16.61%) of scientists in decile 10 did not originate from deciles 8–10, and about 3% came from deciles 1–3 (3.39%). Among all current top-performing late-career scientists in ECON and PSYCH, there are only four Jumpers-Up: two economists (0.52%) and two psychologists (0.36%) who experienced extreme mobility from decile 1 to decile 10 (out of 385 and 548, respectively). For these individuals, we have comprehensive lifetime data on their demographics, publishing and collaboration patterns, and scholarly impact. However, there are no clear patterns regarding who Jumpers-Up actually tend to be and how they tend to publish and collaborate.

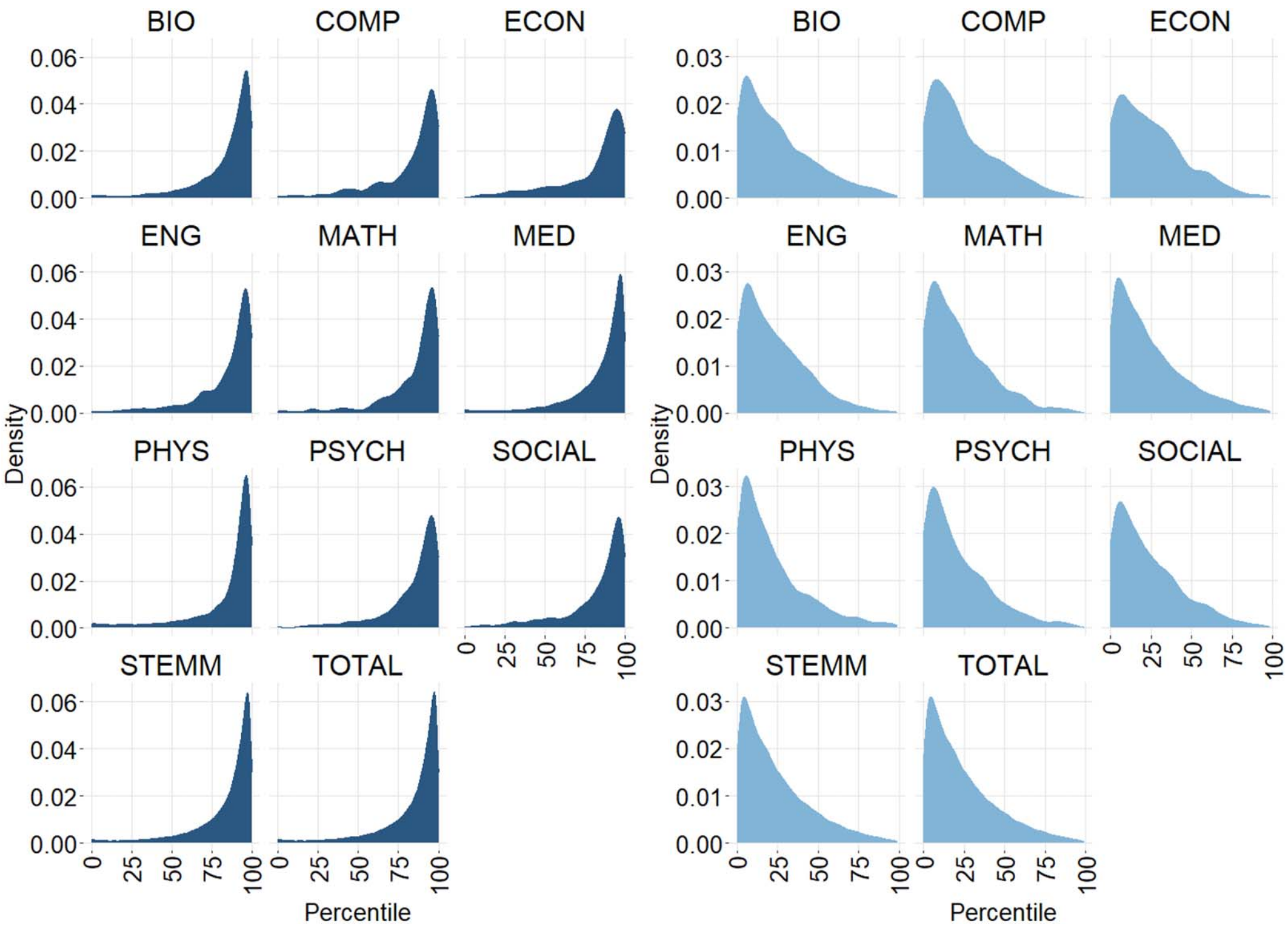

**Figure 4.** (Left panel) How were the top-performing ($N$ = 32,063, productivity decile 10) late-career scientists distributed by the productivity percentiles (range: 0–100) when they were in their midcareer stage? (Right panel) How were the bottom-performing ($N$ = 32,075, productivity decile 1) late-career scientists distributed by productivity percentiles (range: 0–100) when they were in their midcareer stage? Kernel density plots, initial percentile distribution, by selected discipline.

Analogous analyses were performed for the lowest productive scientists (Table 6 in ESM). The observed patterns are similar to, but less pronounced than, those seen for the top performers. Approximately three-fourths (75.31%) of bottom performers originated from the three lowest productivity deciles (deciles 1, 2, and 3), with over one-third coming from the lowest decile (37.41%). Conversely, only 2.28% (728 scientists) came from the top three deciles, including just 0.26% (82 scientists) from the highest decile 10 (referred to as Droppers-Down). Also for these scientists, we have full data (demographics, publishing and collaboration patterns, scholarly impact).

Supplementary Table 7 (in ESM) shows the decile origins of current bottom-performing scientists, revealing similar mobility patterns. The majority of current decile 1 scientists came from the bottom three deciles (68.40%), while only 4.34% originated from the top three deciles. A mere 222 scientists (0.69%) experienced a drop from decile 10 to decile 1

**Table 4.** Mobility of top performers between two career stages: early career (initial stage) and midcareer (target stage): From which initial productivity deciles (at an early-career stage) do top-performing scientists at a midcareer stage come from? Late-career scientists who were top performers at a midcareer stage ($N$ = 32,063) by academic discipline and initial productivity decile (frequencies and percentages)

| | | Total | Bottom 10% | Decile 2 | Decile 3 | Decile 4 | Decile 5 | Decile 6 | Decile 7 | Decile 8 | Decile 9 | Top 10% |
|---|---|---|---|---|---|---|---|---|---|---|---|---|
| **Late-career scientists who were top performers at a midcareer stage** | | | | | | | | | | | | |
| **TOTAL** | ***N*** | **32,063** | **162** | **232** | **323** | **540** | **799** | **1,175** | **2,007** | **3,312** | **6,714** | **16,799** |
| | **%** | **100** | **0.51** | **0.72** | **1.01** | **1.68** | **2.49** | **3.66** | **6.26** | **10.33** | **20.94** | **52.39** |
| **SOCIAL** | ***N*** | **1,259** | **7** | **14** | **12** | **35** | **35** | **48** | **65** | **149** | **234** | **660** |
| | **%** | **100** | **0.56** | **1.11** | **0.95** | **2.78** | **2.78** | **3.81** | **5.16** | **11.83** | **18.59** | **52.42** |
| **STEMM** | ***N*** | **30,804** | **155** | **218** | **311** | **505** | **764** | **1,127** | **1,942** | **3,163** | **6,480** | **16,139** |
| | **%** | **100** | **0.50** | **0.71** | **1.01** | **1.64** | **2.48** | **3.66** | **6.30** | **10.27** | **21.04** | **52.39** |
| AGRI | *N* | 2,373 | 7 | 11 | 9 | 25 | 57 | 72 | 147 | 273 | 510 | 1,262 |
| | % | 100 | 0.29 | 0.46 | 0.38 | 1.05 | 2.40 | 3.03 | 6.19 | 11.50 | 21.49 | 53.18 |
| BIO | *N* | 4,582 | 24 | 41 | 50 | 100 | 121 | 187 | 296 | 476 | 933 | 2,354 |
| | % | 100 | 0.52 | 0.89 | 1.09 | 2.18 | 2.64 | 4.08 | 6.46 | 10.39 | 20.36 | 51.37 |
| BUS | *N* | 326 | 2 | 4 | 2 | 12 | 7 | 16 | 15 | 39 | 66 | 163 |
| | % | 100 | 0.61 | 1.23 | 0.61 | 3.68 | 2.15 | 4.91 | 4.60 | 11.96 | 20.25 | 50.00 |
| CHEM | *N* | 1,490 | 7 | 11 | 14 | 21 | 34 | 50 | 85 | 149 | 306 | 813 |
| | % | 100 | 0.47 | 0.74 | 0.94 | 1.41 | 2.28 | 3.36 | 5.70 | 10.00 | 20.54 | 54.56 |
| COMP | *N* | 765 | 4 | 9 | 20 | 20 | 25 | 29 | 57 | 82 | 150 | 369 |
| | % | 100 | 0.52 | 1.18 | 2.61 | 2.61 | 3.27 | 3.79 | 7.45 | 10.72 | 19.61 | 48.24 |
| EARTH | *N* | 1,437 | 5 | 4 | 14 | 14 | 32 | 51 | 81 | 149 | 297 | 790 |
| | % | 100 | 0.35 | 0.28 | 0.97 | 0.97 | 2.23 | 3.55 | 5.64 | 10.37 | 20.67 | 54.98 |
| ECON | *N* | 385 | 1 | 5 | 5 | 15 | 11 | 13 | 23 | 44 | 64 | 204 |
| | % | 100 | 0.26 | 1.30 | 1.30 | 3.90 | 2.86 | 3.38 | 5.97 | 11.43 | 16.62 | 52.99 |
| ENG | *N* | 1,282 | 7 | 12 | 14 | 26 | 32 | 59 | 88 | 153 | 259 | 632 |
| | % | 100 | 0.55 | 0.94 | 1.09 | 2.03 | 2.50 | 4.60 | 6.86 | 11.93 | 20.20 | 49.30 |
| ENVIR | *N* | 652 | 4 | 2 | 4 | 7 | 17 | 25 | 36 | 64 | 137 | 356 |
| | % | 100 | 0.61 | 0.31 | 0.61 | 1.07 | 2.61 | 3.83 | 5.52 | 9.82 | 21.01 | 54.60 |
| IMMU | *N* | 315 | 1 | 3 | 5 | 7 | 11 | 10 | 26 | 28 | 69 | 155 |
| | % | 100 | 0.32 | 0.95 | 1.59 | 2.22 | 3.49 | 3.17 | 8.25 | 8.89 | 21.90 | 49.21 |
| MATER | *N* | 584 | 3 | 2 | 5 | 11 | 19 | 27 | 38 | 64 | 115 | 300 |
| | % | 100 | 0.51 | 0.34 | 0.86 | 1.88 | 3.25 | 4.62 | 6.51 | 10.96 | 19.69 | 51.37 |
| MATH | *N* | 701 | 3 | 3 | 9 | 11 | 9 | 22 | 37 | 60 | 141 | 406 |
| | % | 100 | 0.43 | 0.43 | 1.28 | 1.57 | 1.28 | 3.14 | 5.28 | 8.56 | 20.11 | 57.92 |
| MED | *N* | 13,108 | 46 | 63 | 113 | 205 | 330 | 491 | 869 | 1,360 | 2,902 | 6,729 |
| | % | 100 | 0.35 | 0.48 | 0.86 | 1.56 | 2.52 | 3.75 | 6.63 | 10.38 | 22.14 | 51.34 |
| NEURO | *N* | 587 | 5 | 6 | 3 | 12 | 16 | 18 | 34 | 51 | 137 | 305 |
| | % | 100 | 0.85 | 1.02 | 0.51 | 2.04 | 2.73 | 3.07 | 5.79 | 8.69 | 23.34 | 51.96 |
| PHYS | *N* | 2,928 | 39 | 51 | 51 | 46 | 61 | 86 | 148 | 254 | 524 | 1,668 |
| | % | 100 | 1.33 | 1.74 | 1.74 | 1.57 | 2.08 | 2.94 | 5.05 | 8.67 | 17.90 | 56.97 |
| PSYCH | *N* | 548 | 4 | 5 | 5 | 8 | 17 | 19 | 27 | 66 | 104 | 293 |
| | % | 100 | 0.73 | 0.91 | 0.91 | 1.46 | 3.10 | 3.47 | 4.93 | 12.04 | 18.98 | 53.47 |

**Table 5.** Mobility of top performers between two career stages: midcareer (initial stage) and late career (target stage): From which initial productivity deciles (at a mid-career stage) do top-performing scientists at a late-career stage come from? Late-career scientists who were top performers at a late-career stage ($N$ = 32,063) by academic discipline and initial productivity decile (frequencies and percentages)

| | | Total | Bottom 10% | Decile 2 | Decile 3 | Decile 4 | Decile 5 | Decile 6 | Decile 7 | Decile 8 | Decile 9 | Top 10% |
|---|---|---|---|---|---|---|---|---|---|---|---|---|
| **Late-career scientists who are top performers at a late-career stage** | | | | | | | | | | | | |
| **TOTAL** | ***N*** | **32,063** | **436** | **305** | **345** | **541** | **756** | **1,125** | **1,816** | **3,139** | **6,339** | **17,261** |
| | **%** | **100** | **1.36** | **0.95** | **1.08** | **1.69** | **2.36** | **3.51** | **5.66** | **9.79** | **19.77** | **53.83** |
| **SOCIAL** | ***N*** | **1,259** | **10** | **14** | **21** | **32** | **44** | **55** | **58** | **131** | **249** | **645** |
| | **%** | **100** | **0.79** | **1.11** | **1.67** | **2.54** | **3.49** | **4.37** | **4.61** | **10.41** | **19.78** | **51.23** |
| **STEMM** | ***N*** | **30,804** | **426** | **291** | **324** | **509** | **712** | **1,070** | **1,758** | **3,008** | **6,090** | **16,616** |
| | **%** | **100** | **1.38** | **0.94** | **1.05** | **1.65** | **2.31** | **3.47** | **5.71** | **9.76** | **19.77** | **53.94** |
| AGRI | *N* | 2,373 | 23 | 16 | 17 | 34 | 49 | 81 | 131 | 245 | 474 | 1,303 |
| | % | 100 | 0.97 | 0.67 | 0.72 | 1.43 | 2.06 | 3.41 | 5.52 | 10.32 | 19.97 | 54.91 |
| BIO | *N* | 4,582 | 65 | 38 | 45 | 85 | 104 | 170 | 266 | 467 | 952 | 2,390 |
| | % | 100 | 1.42 | 0.83 | 0.98 | 1.86 | 2.27 | 3.71 | 5.81 | 10.19 | 20.78 | 52.16 |
| BUS | *N* | 326 | 6 | 3 | 4 | 11 | 14 | 16 | 9 | 38 | 70 | 155 |
| | % | 100 | 1.84 | 0.92 | 1.23 | 3.37 | 4.29 | 4.91 | 2.76 | 11.66 | 21.47 | 47.55 |
| CHEM | *N* | 1,490 | 14 | 7 | 17 | 22 | 32 | 42 | 60 | 133 | 291 | 872 |
| | % | 100 | 0.94 | 0.47 | 1.14 | 1.48 | 2.15 | 2.82 | 4.03 | 8.93 | 19.53 | 58.52 |
| COMP | *N* | 765 | 11 | 8 | 9 | 19 | 32 | 25 | 55 | 59 | 145 | 402 |
| | % | 100 | 1.44 | 1.05 | 1.18 | 2.48 | 4.18 | 3.27 | 7.19 | 7.71 | 18.95 | 52.55 |
| EARTH | *N* | 1,437 | 12 | 21 | 12 | 23 | 42 | 61 | 83 | 155 | 308 | 720 |
| | % | 100 | 0.84 | 1.46 | 0.84 | 1.60 | 2.92 | 4.24 | 5.78 | 10.79 | 21.43 | 50.10 |
| ECON | *N* | 385 | 2 | 8 | 8 | 13 | 14 | 21 | 21 | 31 | 69 | 198 |
| | % | 100 | 0.52 | 2.08 | 2.08 | 3.38 | 3.64 | 5.45 | 5.45 | 8.05 | 17.92 | 51.43 |
| ENG | *N* | 1,282 | 13 | 12 | 19 | 26 | 33 | 43 | 81 | 118 | 240 | 697 |
| | % | 100 | 1.01 | 0.94 | 1.48 | 2.03 | 2.57 | 3.35 | 6.32 | 9.20 | 18.72 | 54.37 |
| ENVIR | *N* | 652 | 8 | 9 | 3 | 11 | 12 | 20 | 47 | 58 | 128 | 356 |
| | % | 100 | 1.23 | 1.38 | 0.46 | 1.69 | 1.84 | 3.07 | 7.21 | 8.90 | 19.63 | 54.60 |
| IMMU | *N* | 315 | 3 | 4 | 5 | 6 | 3 | 16 | 22 | 30 | 61 | 165 |
| | % | 100 | 0.95 | 1.27 | 1.59 | 1.90 | 0.95 | 5.08 | 6.98 | 9.52 | 19.37 | 52.38 |
| MATER | *N* | 584 | 8 | 1 | 6 | 10 | 5 | 15 | 30 | 63 | 117 | 329 |
| | % | 100 | 1.37 | 0.17 | 1.03 | 1.71 | 0.86 | 2.57 | 5.14 | 10.79 | 20.03 | 56.34 |
| MATH | *N* | 701 | 8 | 3 | 12 | 11 | 14 | 14 | 44 | 71 | 123 | 401 |
| | % | 100 | 1.14 | 0.43 | 1.71 | 1.57 | 2.00 | 2.00 | 6.28 | 10.13 | 17.55 | 57.20 |
| MED | *N* | 13,108 | 187 | 131 | 125 | 200 | 312 | 480 | 771 | 1,344 | 2,671 | 6,887 |
| | % | 100 | 1.43 | 1.00 | 0.95 | 1.53 | 2.38 | 3.66 | 5.88 | 10.25 | 20.38 | 52.54 |
| NEURO | *N* | 587 | 9 | 1 | 7 | 6 | 14 | 10 | 33 | 62 | 119 | 326 |
| | % | 100 | 1.53 | 0.17 | 1.19 | 1.02 | 2.39 | 1.70 | 5.62 | 10.56 | 20.27 | 55.54 |
| PHYS | *N* | 2,928 | 65 | 40 | 47 | 56 | 60 | 93 | 135 | 203 | 461 | 1,768 |
| | % | 100 | 2.22 | 1.37 | 1.61 | 1.91 | 2.05 | 3.18 | 4.61 | 6.93 | 15.74 | 60.38 |
| PSYCH | *N* | 548 | 2 | 3 | 9 | 8 | 16 | 18 | 28 | 62 | 110 | 292 |
| | % | 100 | 0.36 | 0.55 | 1.64 | 1.46 | 2.92 | 3.28 | 5.11 | 11.31 | 20.07 | 53.28 |

(Droppers-Down). Upward mobility is of particular interest to science policy; downward mobility, in contrast, may often be attributed to personal life circumstances, such as health issues or family problems, which cannot be fully analyzed through bibliometric data sets.

### 3.3. Changing Productivity Classes: All Academic Disciplines Combined

The Sankey diagram (Figure 5) serves as a visual guide to better help understand the concept of scientists' mobility across productivity classes throughout their careers. This diagram illustrates the movement of scientists between productivity deciles at different career stages: early career (left: top and bottom), midcareer (middle: top and bottom), and late career (right: top and bottom). Our focus is on horizontal top-to-top and bottom-to-bottom mobility as well as the transitions involving extreme downward mobility from top to bottom and extreme upward mobility from bottom to top.

Figure 5 displays the mobility of scientists across all academic disciplines combined ($N$ = 320,564), Figure 6 shows the mobility for all social science disciplines combined ($N$ = 12,585), and Figure 7 shows all STEMM disciplines combined ($N$ = 307,979). The left columns of the diagrams represent the distribution of early-career scientists within the top and bottom productivity deciles (each decile totaling 100%), the middle columns represent midcareer scientists, and the right columns represent late-career scientists within the same two productivity classes. To enhance clarity, deciles 2 through 9 are excluded from the diagram.

The horizontal top-to-top and bottom-to-bottom mobility between the early- and midcareer stages is represented by thick flows: More than a half of global top performers continue as global top performers (52.39%), and more than one-third of global bottom performers continue as global bottom performers (37.41%). Extreme vertical top-to-bottom and bottom-to-top mobilities are rare and represented as thin downward and upward flows: Only 0.26% of top productive early-career scientists (82 scientists) land in the bottom productivity midcareer scientists—and only 0.51% bottom productive early-career scientists (162 scientists) land in the class of top productive midcareer scientists.

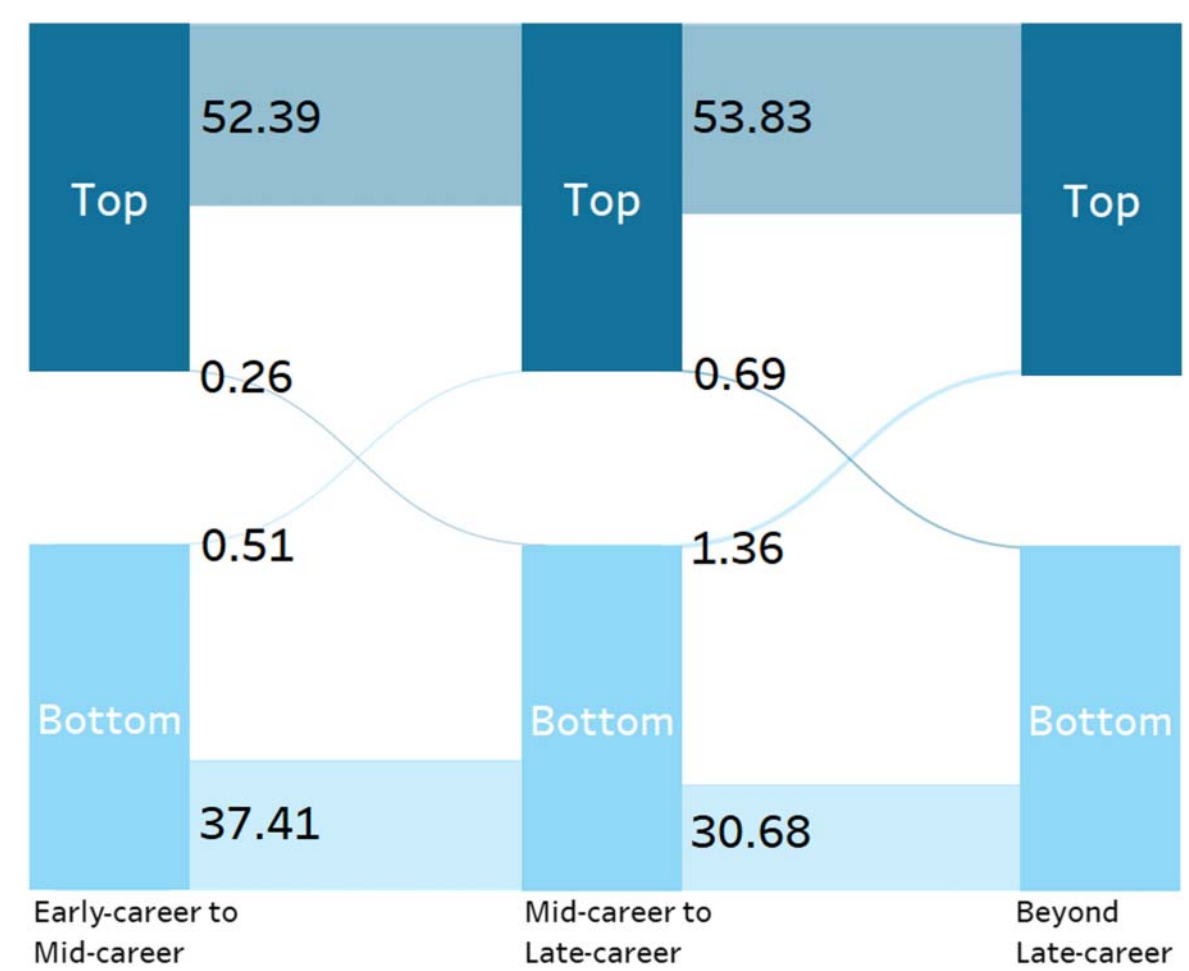

**Figure 5.** Sankey diagram of retrospectively constructed mobility between productivity classes in the three stages of a scientific career. All academic disciplines combined (TOTAL), current late-career scientists. All observations ranked and clustered into productivity deciles, top- (upper 10%, productivity decile 10) and bottom- (bottom 10%, productivity decile 1) productivity classes only, other productivity deciles removed for clarity ($N$ = 320,564) (percentages, top class, and bottom class, 100% each).

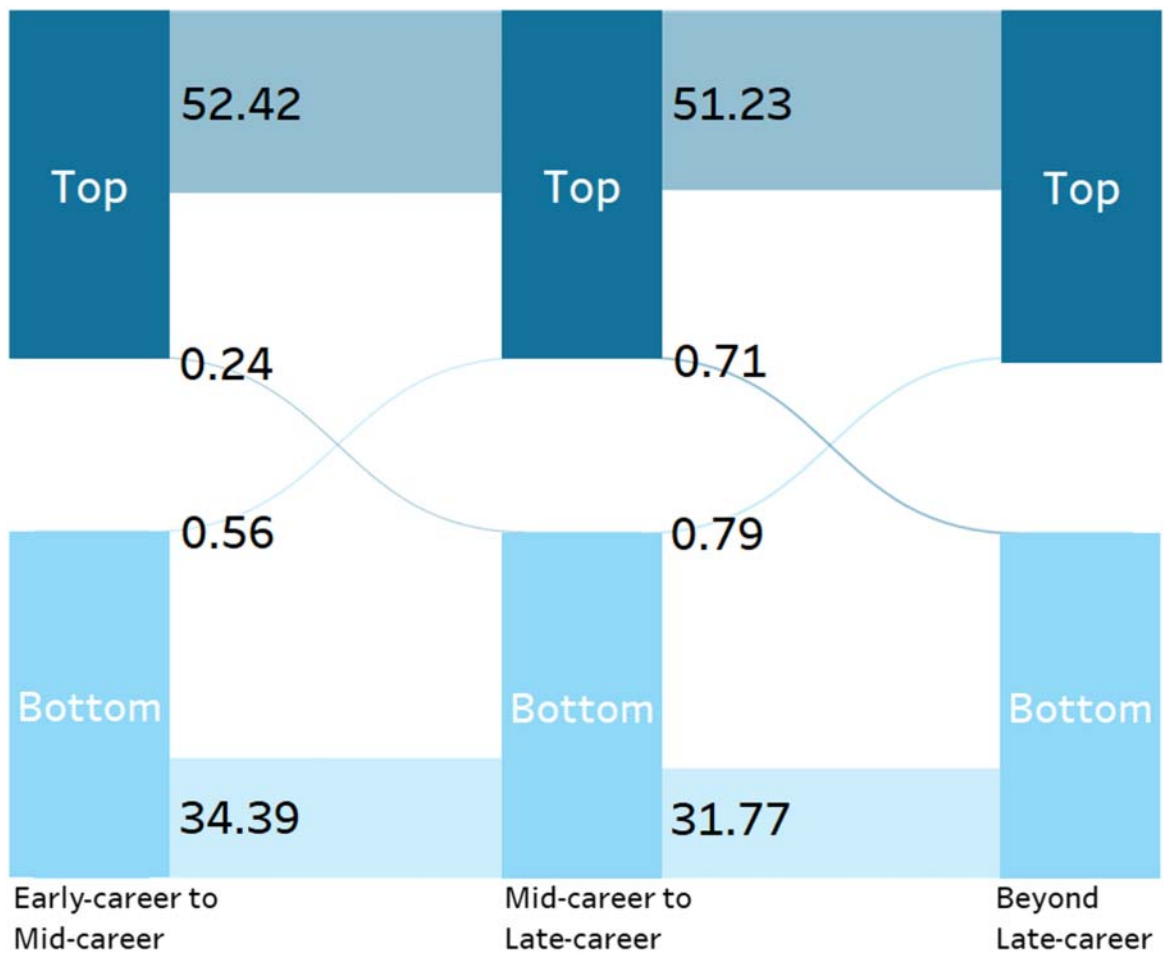

**Figure 6.** Sankey diagram of retrospectively constructed mobility between productivity classes in the three stages of a scientific career. All social science disciplines combined (SOCIAL), current late-career scientists. All observations ranked and clustered into productivity deciles, top (upper 10%, productivity decile 10) and bottom (bottom 10%, productivity decile 1) productivity classes only, other productivity deciles removed for clarity (*N* = 12,585) (percentages, top class and bottom class, 100% each).

The mobility patterns between midcareer and late-career stage are very similar. The mobility patterns do not differ between social science disciplines combined (Figure 6) and STEMM disciplines combined (Figure 7): Surprisingly, despite different publishing and collaboration patterns, the top-to-top mobility for the early- to midcareer transition is almost exactly the same; and for the mid- to late-career transition it is slightly higher for STEMM disciplines.

It is as rare in the SOCIAL disciplines to experience extreme upward mobility—from decile 1 to decile 10—as it is in the STEMM disciplines. In our sample of SOCIAL scientists (*N* = 12,585), there are only seven scientists involved in the first transition and only 10 scientists involved in the second transition out of 1,259 scientists (Table 6). Here and elsewhere, the statistical significances of differences are not shown because we work with a population of scientists (all scientists meeting the inclusion criteria) rather than with their sample (a selection of all scientists), which is an essential difference between small-scale and large-scale studies.

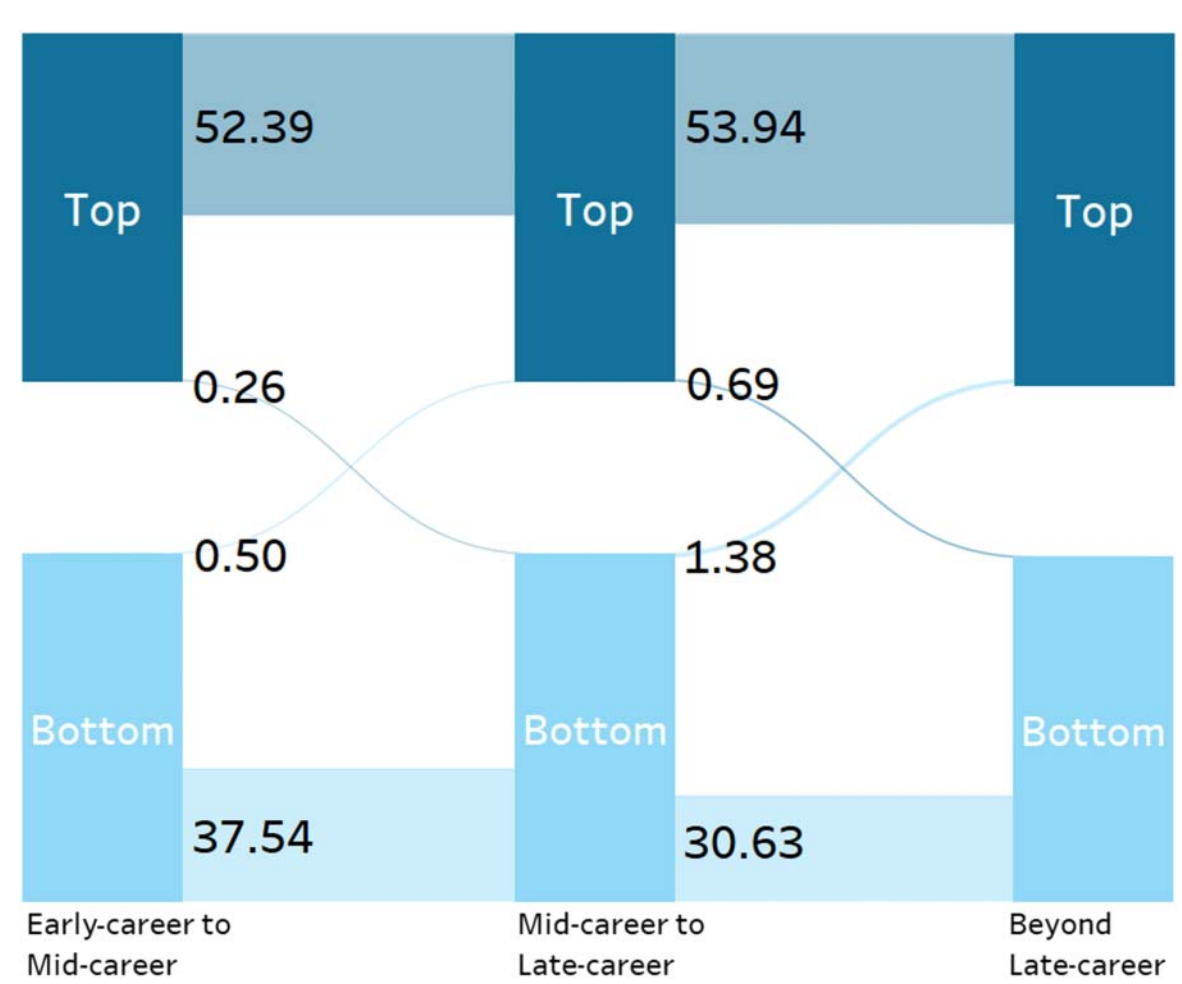

**Figure 7.** Sankey diagram of retrospectively constructed mobility between productivity classes in the three stages of a scientific career. All STEMM academic disciplines combined, current late-career scientists. All observations ranked and clustered into productivity deciles, top (upper 10%, productivity decile 10) and bottom (bottom 10%, productivity decile 1) productivity classes only, other productivity deciles removed for clarity (*N* = 307,979) (percentages, top class and bottom class, 100% each).

From an aggregated perspective of all academic disciplines combined (Table 6), the mobility patterns are clear: Over half (52.39% and 53.83%) of the scientists who achieve top productivity (decile 10) remain in this top category. Similarly, about one-third (37.41% and 30.68%) of those in the bottom productivity category (decile 1) continue to stay in the same class. This indicates an intriguing "locking-in" mechanism within academic careers that warrants further scholarly investigation.

The likelihood of experiencing extreme upward mobility (moving from decile 1 to decile 10) or extreme downward mobility (moving from decile 10 to decile 1) between productivity classes is very low. According to our data (Table 6), in the context of our prestige-normalized counting approach, the chances of radical change in publishing behavior compared with peers within an academic discipline are minimal.

Specifically, only 162 scientists (0.51%) who became top performers in their midcareer stage were initially in the bottom productivity class during their early-career stage. Similarly, only 436 scientists (1.36%) who were top performers in their late-career stage started as bottom performers in their midcareer stage. The chances of extreme downward mobility are also rare, with less than 1% of scientists moving from the top to the bottom productivity class in successive career stages (0.26% and 0.69%, respectively). These data indicate that radical changes in publishing behavior that lead to such significant shifts in productivity classes are quite uncommon.

### 3.4. Changing Productivity Classes: Cross-Disciplinary Differentiation

The aggregated pictures of all academic disciplines combined, all STEMM disciplines combined, and all SOCIAL disciplines combined hide a much more nuanced picture of individual academic disciplines with their distinct mobility patterns between productivity classes.

Focusing on the horizontal top-to-top (productivity decile 10 to decile 10) mobility first, for almost all academic disciplines, more than 50% of the top productivity scientists continue as top productivity scientists (Table 7). The highest share is observed for MATH and PHYS in both transitions, reaching as much as 60.38% for PHYS in the second transition. Scientists representing bottom-to-top mobility (Jumpers-Up), which is of great interest in productivity studies, are extremely rare across all academic disciplines: Their share ranges from 0.26% for ECON to 1.33% for PHYS in the first career transition and from 0.36% for PSYCH and 2.22% for BUS in the second career transition.

Specifically, in MED (the largest discipline), more than 50% of the top productivity scientists continue as top productivity scientists (51.34% and 52.54%) and about one-third of bottom productivity scientists continue as bottom productivity scientists (36.37% and 29.48%)—which is about the average for all STEMM disciplines. Interestingly, top-to-bottom mobility in the early-career stage in MED is one of the lowest among disciplines examined.

In the two math-intensive disciplines (MATH and PHYS), persistence in high productivity is the highest for both career stages, while for the other two from the math-intensive cluster (COMP and ENG), it is among the lowest for the first career stage. Interestingly, the patterns for social science disciplines follow the patterns for STEMM disciplines: The mechanisms of

**Table 6.** Mobility between top (decile 10) and bottom (decile 1) productivity classes while moving up from the early-career stage to Midcareer stage and from the midcareer to late-career stage by all academic disciplines combined ($N$ = 320,564, top panel), social science disciplines combined ($N$ = 12,585, middle panel), and STEMM disciplines combined ($N$ = 307,979, bottom panel) (frequencies and percentages)

| Career stage (transition from) | Initial productivity decile | Career stage (transition to) | Target productivity decile | Number of scientists in transition | Number of scientists in productivity class | % |
|---|---|---|---|---|---|---|
| ***Total (all disciplines combined)*** | | | | | | |
| Early career | Bottom | Midcareer | Bottom | 11,996 | 32,063 | 37.41 |
| Early career | Bottom | Midcareer | Top | 162 | 32,063 | 0.51 |
| Early career | Top | Midcareer | Bottom | 82 | 32,063 | 0.26 |
| Early career | Top | Midcareer | Top | 16,799 | 32,063 | 52.39 |
| Midcareer | Bottom | Late career | Bottom | 9,836 | 32,063 | 30.68 |
| Midcareer | Bottom | Late career | Top | 436 | 32,063 | 1.36 |
| Midcareer | Top | Late career | Bottom | 222 | 32,063 | 0.69 |
| Midcareer | Top | Late career | Top | 17,261 | 32,063 | 53.83 |
| ***All social science disciplines*** | | | | | | |
| Early career | Bottom | Midcareer | Bottom | 433 | 1,259 | 34.39 |
| Early career | Bottom | Midcareer | Top | 7 | 1,259 | 0.56 |
| Early career | Top | Midcareer | Bottom | 3 | 1,259 | 0.24 |
| Early career | Top | Midcareer | Top | 660 | 1,259 | 52.42 |
| Midcareer | Bottom | Late career | Bottom | 400 | 1,259 | 31.77 |
| Midcareer | Bottom | Late career | Top | 10 | 1,259 | 0.79 |
| Midcareer | Top | Late career | Bottom | 9 | 1,259 | 0.71 |
| Midcareer | Top | Late career | Top | 645 | 1,259 | 51.23 |
| ***All STEMM disciplines*** | | | | | | |
| Early career | Bottom | Midcareer | Bottom | 11,563 | 30,804 | 37.54 |
| Early career | Bottom | Midcareer | Top | 155 | 30,804 | 0.50 |
| Early career | Top | Midcareer | Bottom | 79 | 30,804 | 0.26 |
| Early career | Top | Midcareer | Top | 16,139 | 30,804 | 52.39 |
| Midcareer | Bottom | Late career | Bottom | 9,436 | 30,804 | 30.63 |
| Midcareer | Bottom | Late career | Top | 426 | 30,804 | 1.38 |
| Midcareer | Top | Late career | Bottom | 213 | 30,804 | 0.69 |
| Midcareer | Top | Late career | Top | 16,616 | 30,804 | 53.94 |

**Table 7.** Four mobility types by academic discipline between the early-career and midcareer stage, and between the midcareer and late-career stage, percentages ($N$ = 320,564)

| Academic discipline | Top-to-top mobility | | Bottom-to-bottom mobility | | Top-to-bottom mobility | | Bottom-to-top mobility | |
|---|---|---|---|---|---|---|---|---|
| | Early-career stage (%) | Midcareer stage (%) | Early-career stage (%) | Midcareer stage (%) | Early-career stage (%) | Midcareer stage (%) | Early-career stage (%) | Midcareer stage (%) |
| AGRI | 53.18 | 54.91 | 38.60 | 32.20 | 0.17 | 0.25 | 0.29 | 0.97 |
| BIO | 51.37 | 52.16 | 36.56 | 28.26 | 0.33 | 0.87 | 0.52 | 1.42 |
| BUS | 50.00 | 47.55 | 30.37 | 26.38 | 0.31 | 0.61 | 0.61 | 1.84 |
| CHEM | 54.56 | 58.52 | 41.01 | 33.42 | 0.07 | 0.40 | 0.47 | 0.94 |
| COMP | 48.24 | 52.55 | 32.42 | 28.37 | 0.13 | 0.26 | 0.52 | 1.44 |
| EARTH | 54.98 | 50.10 | 40.36 | 35.00 | 0.21 | 0.28 | 0.35 | 0.84 |
| ECON | 52.99 | 51.43 | 29.87 | 29.09 | 0.26 | 0.78 | 0.26 | 0.52 |
| ENG | 49.30 | 54.37 | 36.97 | 29.88 | 0.23 | 0.47 | 0.55 | 1.01 |
| ENVIR | 54.60 | 54.60 | 35.58 | 28.68 | 0.46 | 0.46 | 0.61 | 1.23 |
| IMMU | 49.21 | 52.38 | 35.87 | 35.87 | 0.63 | – | 0.32 | 0.95 |
| MATER | 51.37 | 56.34 | 38.70 | 32.53 | – | 0.34 | 0.51 | 1.37 |
| MATH | 57.92 | 57.20 | 36.23 | 32.38 | 0.29 | 0.29 | 0.43 | 1.14 |
| MED | 51.34 | 52.54 | 36.37 | 29.48 | 0.16 | 0.82 | 0.35 | 1.43 |
| NEURO | 51.96 | 55.54 | 39.18 | 31.35 | 0.34 | 0.34 | 0.85 | 1.53 |
| PHYS | 56.97 | 60.38 | 42.25 | 34.53 | 0.75 | 1.13 | 1.33 | 2.22 |
| PSYCH | 53.47 | 53.28 | 39.96 | 36.86 | 0.18 | 0.73 | 0.73 | 0.36 |
| **SOCIAL** | **52.42** | **51.23** | **34.39** | **31.77** | **0.24** | **0.71** | **0.56** | **0.79** |
| **STEMM** | **52.39** | **53.94** | **37.54** | **30.63** | **0.26** | **0.69** | **0.50** | **1.38** |
| **TOTAL** | **52.39** | **53.83** | **37.41** | **30.68** | **0.26** | **0.69** | **0.51** | **1.36** |

*Note*: "–" = no scientists involved in this transition.

survival among high performers seem exactly the same in both fields usually treated separately in science policy studies.

### 3.5. Model Approach: Logistic Regression

In this section, we introduce a multidimensional approach and analyze the odds ratio estimates of membership in the classes of global top and bottom productive scientists for current late-career scientists and, retrospectively, for current late-career scientists when they were midcareer scientists (the upper 10% and the bottom 10%, or decile 10 and decile 1, with separate models for each discipline, $N$ = 320,564).

We use a single demographic variable (gender, binary: male or female) and four variables we have computed using microlevel data about individual scientists. One variable is related to the article-level metrics of individual scholarly impact (field-normalized citations received within the first 4 years after each article has been published); two other variables are related to individual publishing and collaboration patterns (lifetime median team size and lifetime international collaboration rate); and one variable is related to publishing productivity in earlier career stages (prior membership in the global top and bottom productivity classes at the early-career stage and at the midcareer stage).

All publications (lifetime) and cited references (lifetime) were used to compute a unique discipline to which every scientist was ascribed (see Dataflow in Figure 1). The field-weighted 4-year citation impact is computed for each individual publication separately and subsequently averaged to all publications over one's lifetime; publishing and collaboration pattern variables are also computed from a lifetime perspective of individual scientists: All journal articles and articles in conference proceedings published throughout one's lifetime are examined. In contrast, membership in productivity classes has been computed for two specific periods of careers (early- and midcareer periods). In addition, we also use in our regression models an institutional variable (TOP 200 institution globally, for logistic regression analysis for top productive late-career scientists only). The variables and their short descriptions are presented in Table 1. The inverse correlation matrixes and main diagonals are shown in Supplementary Tables 8–10 in ESM.

Extensive previous research on individual productivity has suggested that the most important predictors of high research productivity at the individual level are international collaboration (Dusdal & Powell, 2021), working in teams (Wagner, 2018), gender (Fox & Nikivinicze, 2021; Larivière et al., 2013), career stage and academic experience (Jung, 2014; Shin & Cummings, 2010; Kwiek, 2016), and publication productivity earlier in academic careers (Horta & Santos, 2016; Kwiek & Roszka, 2024a, 2024b), consistent with the Matthew effect in science (the rich get disproportionately richer while the poor get poorer; DiPrete & Eirich, 2006).

As a result—and specifically in the context of our two-dimensional results visualized through Sankey diagrams that show powerful top-to-top and bottom-to-bottom mobility in terms of productivity—we have added the prior membership in global top and bottom productivity classes as variables used in the regression analysis.

First, we analyze the top productivity classes: odds ratio estimates of membership in the class of global top productive midcareer scientists (Table 8) and global top productive late-career scientists (Table 9). In the vast majority of disciplines, high productivity in an earlier career stage is the most powerful predictor of high productivity in a later stage, with Exp(B) in the range of 11.68 (IMMU) to 22.97 (MATH) for the first career stage and 13.89 (EARTH) to 21.82 (MATH) for the second career stage (in all cases: all other things being equal; the predictor is not statistically significant for two disciplines in the first stage and six disciplines in the second stage).

The direction of the impact of the gender variable (being a male scientist) is consistent across disciplines; however, the impact is highly differentiated. Being male increases the probability of success in 13 out of 16 disciplines for midcareer scientists by as little as 3–7% in COMP, MATH, and ECON and by more than 100% (in IMMU and NEURO). For late-career scientists, gender is statistically significant in 10 disciplines, and its impact is smaller; being male increases the odds of success by 3–14% (in AGRI, EARTH, MATER) to 46–70% (in NEURO, BIO, and IMMU), with the exception of MATH, where it decreases the odds by 2%; again, all other things being equal.

Interestingly, the probability of success for men differs between highly mathematized disciplines, which traditionally have the lowest percentage of women (MATH, COMP, ENG,

**Table 8.** Logistic regression statistics: odds ratio estimates of membership in the class of global top productive midcareer scientists (the top 10%, separately for each academic discipline) (*N* = 320,564)

| Variables | AGRI | BIO | BUS | CHEM | COMP | EARTH | ECON | ENG |
|---|---|---|---|---|---|---|---|---|
| $R^2$ | 0.25 | 0.22 | 0.21 | 0.26 | 0.18 | 0.26 | 0.23 | 0.20 |
| Male | 1.32*** | 1.86*** | 1.61*** | 1.30*** | 1.07*** | 1.29*** | 1.03* | 0.78 |
| Avg. FWCI 4y Early | 1.53*** | 1.14*** | 1.04*** | 1.37*** | 1.05*** | 1.24*** | 1.12*** | 1.16*** |
| Inter. Collab. Rate Early | 1.01*** | 1.01*** | 1.01*** | 1.01*** | 1.00*** | 1.01*** | 1.01*** | 1.01*** |
| Median Team Size Early | 1.06*** | 1.01*** | 1.31*** | 0.99*** | 1.12*** | 1.05*** | 1.35*** | 0.96** |
| Top Early | 15.39*** | 15.87*** | 15.10*** | 17.81*** | 14.64*** | 17.17*** | 17.49*** | 14.33 |
| Constant | 0.02*** | 0.02*** | 0.02* | 0.02* | 0.03* | 0.02** | 0.02* | 0.06 |
| | **ENVIR** | **IMMU** | **MATER** | **MATH** | **MED** | **NEURO** | **PHYS** | **PSYCH** |
| $R^2$ | 0.26 | 0.21 | 0.24 | 0.28 | 0.22 | 0.25 | 0.34 | 0.25 |
| Male | 1.21 | 2.12*** | 1.41* | 1.05* | 1.30*** | 2.27*** | 0.95 | 1.36** |
| Avg. FWCI 4y Early | 1.27*** | 1.22*** | 1.41*** | 1.23*** | 1.02*** | 1.37*** | 1.05*** | 1.08*** |
| Inter. Collab. Rate Early | 1.01*** | 1.01*** | 1.01*** | 1.00*** | 1.01*** | 1.00*** | 1.02*** | 1.01*** |
| Median Team Size Early | 1.01*** | 1.02*** | 1.06*** | 1.40*** | 1.08*** | 1.16*** | 1.17*** | 1.25*** |
| Top Early | 18.16 | 11.68*** | 13.53** | 22.97*** | 16.39*** | 14.25*** | 11.92** | 18.52** |
| Constant | 0.02 | 0.02** | 0.02 | 0.02 | 0.03*** | 0.01** | 0.01 | 0.02 |

* $p < 0.05$.

** $p < 0.01$.

*** $p < 0.001$.

and PHYS), and the disciplines with high percentages of women (see Table 1 in ESM for male and female participation data in our sample). In the former cluster, the impact of gender is marginal (COMP and MATH in the first stage; MATH in the second stage) or not statistically significant (ENG and PHYS in the first stage, and COMP, ENG, and PHYS in the second stage).

In contrast, in the disciplines with high participation of women, the probability of success for men is 86% higher than for women in BIO, 112% higher in IMMU, and 127% higher in NEURO (for the first stage), and it is 70% higher in IMMU, 46% higher in NEURO, and 47% higher in BIO, the largest exception to the rule being MED, which has much lower increases for men compared with women.

For all disciplines, at both career stages, article-level citation metrics are statistically significant (except for PSYCH and COMP in the second stage): An increase of a field-weighted citation impact for all publications from an early-career period by one unit increases the probability of success from a few percentage points (as in BUS and PSYCH, as well as COMP and PHYS) to as much as 53% in AGRI, all other things being equal. In the second stage, the impact in the midcareer stage is the most consequential in MATER, increasing the odds by 50%.

**Table 9.** Logistic regression statistics: odds ratio estimates of membership in the class of global top productive late-career scientists (the top 10%, separately for each academic discipline) (*N* = 320,564)

| Variables | AGRI | BIO | BUS | CHEM | COMP | EARTH | ECON | ENG |
|---|---|---|---|---|---|---|---|---|
| $R^2$ | 0.25 | 0.22 | 0.19 | 0.29 | 0.22 | 0.21 | 0.21 | 0.26 |
| Male | 1.07*** | 1.47*** | 1.18 | 1.27*** | 0.97 | 1.14** | 1.26*** | 0.89 |
| Avg. FWCI 4y Mid | 1.31*** | 1.03*** | 1.12*** | 1.28*** | 1.05 | 1.10*** | 1.03*** | 1.09** |
| Inter. Collab. Rate Mid | 1.01*** | 1.01*** | 1.01*** | 1.01*** | 1.01*** | 1.01*** | 1.00*** | 1.02*** |
| Median Team Size Mid | 0.99*** | 1.03*** | 1.17 | 0.93*** | 1.05 | 1.00*** | 1.34*** | 0.89* |
| TOP200 | 1.46*** | 1.30*** | 1.26 | 1.42*** | 1.09 | 1.47*** | 1.01*** | 1.81 |
| Top Mid | 18.06*** | 16.76*** | 12.47 | 22.98*** | 17.92 | 13.89*** | 16.16*** | 18.01 |
| Constant | 0.03* | 0.03*** | 0.02 | 0.03(1) | 0.04 | 0.03(1) | 0.02(1) | 0.05 |
| | **ENVIR** | **IMMU** | **MATER** | **MATH** | **MED** | **NEURO** | **PHYS** | **PSYCH** |
| $R^2$ | 0.25 | 0.24 | 0.28 | 0.28 | 0.23 | 0.26 | 0.35 | 0.24 |
| Male | 1.09 | 1.70** | 1.13* | 0.98* | 1.24*** | 1.46*** | 0.85 | 0.95 |
| Avg. FWCI 4y Mid | 1.18*** | 1.14*** | 1.50*** | 1.28*** | 1.02*** | 1.20*** | 1.07*** | 1.33 |
| Inter. Collab. Rate Mid | 1.01*** | 1.01*** | 1.01*** | 1.00*** | 1.01*** | 1.01*** | 1.02*** | 1.01*** |
| Median Team Size Mid | 0.92*** | 1.04*** | 0.95*** | 1.35*** | 1.10*** | 1.02*** | 1.10*** | 1.04 |
| TOP200 | 1.51 | 1.34* | 1.44* | 1.05** | 1.50*** | 1.41*** | 1.23 | 1.43 |
| Top Mid | 18.83 | 14.73** | 18.26*** | 21.82*** | 16.47*** | 19.27*** | 13.94 | 16.74 |
| Constant | 0.04 | 0.02 | 0.02 | 0.02 | 0.02*** | 0.02** | 0.01 | 0.03 |

* $p < 0.05$.

** $p < 0.01$.

*** $p < 0.001$.

In addition, international collaboration rate at an early-career period (and at a midcareer period) is statistically significant for all disciplines, from both the STEMM and SOCIAL clusters of academic disciplines. Generally, the probability of success increases by about 1% for a one-unit increase so that a 30% higher rate increases the odds by 30%.

Finally, team size is statistically significant: The higher the median team size at an early-career period, the higher chances of success. The team size is especially consequential for membership in top productive midcareer class in traditionally sole-authored or small-team disciplines, such as the three social science disciplines (BUS, ECON, and PSYCH) and MATH in the STEMM cluster. In these disciplines, a one-unit increase (i.e., one more coauthor in publications in an early-career period) increases the probability of success on average by 25–40%. In other words, publishing on average with three additional coauthors—or working in larger teams—increases the probability of success by as much as 75–120%.

However, at a late career stage, the role of larger collaboration teams is much more diversified: In some academic disciplines, larger median team size at a midcareer period actually decrease the chances of success; in several disciplines, the team size is statistically insignificant (including BUS, COMP, and PSYCH); and in others, team size continues to be a highly influential predictor of success (ECON and MATH where a one-unit increase increases the odds by as much as one-third).

In models for the late-career stage, we have introduced a TOP200 variable (see Table 1). The TOP200 predictor substantially (by 30–50%) increases the odds of success in eight disciplines, with the highest impact in MED, the largest discipline (by 50%—which may be linked to better access to better training and costly infrastructure such as university hospitals). The predictor is statistically insignificant in such disciplines as BUS and PSYCH as well as COMP and PHYS.

We have also analyzed the global bottom productivity classes (Supplementary Tables 12–13 in ESM). The results are mirror images of those for global top productivity classes but only to some extent. Specifically, although the role of membership in the bottom productivity classes in previous career stages follows the expected pattern—that is, prior membership increases the odds of future membership—the independent variables of prior bottom productivity classes are statistically significant for only seven (first career period) and two (second career period) disciplines. Membership in the bottom productivity classes earlier (publishing years 5–14) increases the probability of membership in this class later (publishing years 15–24) by four to seven times in the case of the first career period and by three times (BIO and MED only, the two largest disciplines in our sample, comprising together 55.18% scientists) in the case of the second career period.

The role of gender is not unequivocal: Being a female scientist substantially increases the odds of membership in global bottom productivity classes only in AGRI and ENG in the first period and slightly (by 17%) in MED in the second stage; interestingly, for all disciplines but two (MED and BIO), gender is statistically insignificant in the second stage.

What is notable is the contrast between BIO (with one of the highest shares of late-career female scientists) and ENG (with the lowest share of late-career female scientists) in the case of midcareer scientists. Being male in BIO decreases the odds of success by one-third on average, but being male in ENG increases the odds of success by half on average (28.0% and 48.0%, respectively).

### 3.6. Limitations

The present research represents trade-offs between what is theoretically desirable and what is practically possible in studying the global patterns of individual research productivity, here based on currently available data from global bibliometric data sets. However, there are trade-offs, biases, and limitations related to our data and methodology.

First, there seem to be no longitudinal data sets globally available (for 38 OECD countries), other than Scopus (or Web of Science—but not OpenAlex; see Priem, Piwowar, & Orr, 2022), that can be meaningfully used to examine changing field-normalized productivity over scientists' lifetimes. In terms of author disambiguation, Scopus is more accurate than Web of Science (Sugimoto & Larivière, 2018, p. 36), but it is certainly not perfect. National data sets are available for selected countries only and with selected parameters only (e.g., the CRISTIN data set for Norway, the Academic Analytics data set for the United States, and the RADON data set for Poland; see Albarrán, Crespo et al., 2011; Kwiek & Roszka, 2024b; Nygaard, Aksnes, & Piro, 2022; Savage & Olejniczak, 2021). As a result, no longitudinal and discipline-based global (as opposed to selected country-based) approaches to publishing productivity are currently possible without access to the global bibliometric data sets that provide metadata on all indexed publications over time. However, global bibliometric data sets come with their own limitations and biases, as has been widely discussed for at least two decades (see, e.g., Sugimoto & Larivière, 2018, pp. 38–44, on the cultural biases of bibliometric data sources; Sugimoto & Larivière, 2023, pp. 11–12, on published scientific document standing as a proxy for the complex interpersonal and cognition processes of knowledge production that precede it).

Second, the character of our data set determines a reductive understanding of individual productivity in which only Scopus-indexed publications are counted, leaving aside nonindexed publications in English and most publications in local languages. However, our focus on STEMM and the three selected social science disciplines, here generally using English for global scholarly communication, makes the present research less biased (STEMM disciplines are covered in global data sets in much larger percentages than the humanities). Scopus is reported as being the largest quality-controlled citation index, covering substantially more years than Dimensions or the Web of Science Core Collection (Thelwall & Sud, 2022). Additionally, all nonpublishing academic activities do not count toward productivity, and the nonpublishers in science have not been analyzed.

Third, the longitudinal nature of our study makes only the survivors our focus: We leave aside all scientists who are not research active for at least 25 years, which follows our definition of late-career scientists. As a result, being aware of high attrition rates in the STEMM disciplines (in OECD countries as recently analyzed in Kwiek & Szymula, 2024, 2025; and in the United States as recently analyzed in Spoon, LaBerge et al., 2023), we acknowledge a “success bias” in our research: The various mobility types between the analyzed productivity classes do not actually refer to all beginning, early-career, and midcareer scientists (active in publishing for less than five, less than 15, and less than 25 years, respectively) but to survivors in science only (see Wang & Barábasi, 2021, pp. 241–245 on the underexploration of “failures” in science). Our study takes a long-term view in which, necessarily, because of the high attrition rate in science, the majority of currently active scientists are actually not represented.

Fourth, our methodology has clear limitations that are especially evident if we compare the present study to single-nation longitudinal studies of publishing productivity. In single-nation studies, a wealth of national data are used (e.g., individual career histories with academic promotion dates, doctoral and postdoctoral dissertation details, research funding details, national classifications of disciplines, national rankings of institutions) that are currently not available for the 38 OECD countries. Additionally, our global study examines scientists from systems with different research funding levels and average individual publishing productivity.

Therefore, out of necessity, our analyses have had to rely on several proxies:

- a commercial journal classification (Scopus All Science Journal Classification system, ASJC) and journal disciplinary classifications rather than a wealth of national disciplinary classifications (Baas, Schotten et al., 2020);
- data on individual Scopus IDs rather than data on “real scientists” with their national registry-based IDs (creating a fundamental ontological difference between traditional academic career research and bibliometric and bibliometric data-driven studies);
- inferred rather than self-declared (or administratively-provided) gender, here based on gender-determining algorithms (probability threshold used: 0.85);
- a single country affiliation and a single institutional affiliation rather than a plethora of changing country and institutional affiliations over a course of academic careers, at least for some percentage of scientists;

- academic careers beginning with the first indexed publication and 25 years of research experience counted based on this publication date (i.e., publishing career)—rather than first employment in academia or beyond (i.e., academic or scientific career *per se*). In our research, the breadth of scientists' activities in academia (e.g., mentoring students, refereeing papers, reviewing grant proposals, and editing journals) is ignored (Liu et al., 2023).

Fifth, our approach to publishing productivity uses journal-level metrics of impact rather than more granular article-level metrics of impact, with all limitations (as discussed, e.g., in Weingart (2004); an improvement of the present research would be to implement productivity measures based on the citation impact directly achieved by each individual publication, as suggested but not implemented in Carrasco and Ruiz-Castillo (2014); however, in logistic regression analyses, we use the field-weighted 4-year citation impact assigned to each scientist and based on averaged citations to every article, here using the Scopus 4-digit ASJC discipline classification).

Sixth, our data set does not include environmental variables that bear heavily on individual publishing productivity, especially on the productivity of women scientists: We were not able to examine "work climates" characterizing the basic units in which scientists work, which are reported to be especially important in STEM disciplines for both attrition (Spoon et al., 2023) and productivity (Branch, 2016; Fox & Mohapatra, 2007). We had no data about academic attitudes and behaviors, working time distribution, teaching/research preference, work-life balance, and household and parenting obligations, which have been routinely reported in productivity studies based on survey data (e.g., Kwiek (2016) on 11 European countries and Kwiek (2018) on Poland), which leads us to the final limitation of the current study: no gender differentiation in examining publishing careers.

Seventh, productivity is one measure of scientists' success, and scholarly impact in the long term is another. Citations tell as much about individual academic success as publications and publishing patterns and intensity. The focus in the present paper was on changing productivity classes rather than changing impact classes (which would also be possible to do based on individual-level, field-weighted citation impact (FWCI) of all articles published within specific career periods). However, our reference to impact based on citations was twofold: through Scopus journal citation ranks (range: 0–100) in productivity computations and through the variable of FWCI 4y in regression models (or the impact of every publication within 4-year windows).

Finally, our research does not tackle an extremely relevant topic of gender differences in mobility between productivity classes over time, except for the regression models. Adding an additional dimension of gender to an examination of late-career scientists—technically perfectly feasible, with full data available—would complicate our global mobility picture substantially; as a result, we have decided to leave a study of gender disparities by discipline for another occasion to make sure this topic can be reflected on with sufficient depth.

## 4. DISCUSSION AND CONCLUSIONS

Our global results fully support our previous small-scale single-nation research, in which half of the Polish top productive assistant professors continued as top productive associate professors, and half of the top productive associate professors continued as top productive full professors (52.6% and 50.8%) (we used a much less granular approach: a tripartite division into top, middle, and bottom productivity classes based on the 20/60/20 formula, $N$ = 2,326 full professors; Kwiek & Roszka, 2024a). Similarly, publishing productivity during assistant professorship heavily influences productivity during associate professorship ($N$ = 4,165 associate professors; Kwiek & Roszka, 2024b). Individual-level microdata from a national registry of scientists suggested that Polish associate professors tend to be stuck in their productivity classes for years: High performers tend to remain high performers, and low performers tend to remain low performers for up to 40 years of their careers.

Our focus is not on globally evolving productivity over time (Rørstad & Aknes, 2015) or evolving productivity from a generational perspective (e.g., the old in science being more productive than the young or the other way round: Savage & Olejniczak, 2021; Way et al., 2017) but on productivity interclass mobility of individuals over their entire scientific careers.

We suggest that, relatively early on in scientific careers, the productivity distribution within the global science profession at its two extremes (top 10% and bottom 10%) is already largely settled and that the early global distribution persists over time, that is, for years and decades. Exceptions are very rare: global bottom performers almost never become global top performers (our Jumpers-Up), and global top performers almost never become global bottom performers (our Droppers-Down).

However, global data on individual publishing careers show that attrition in science—leaving science for good or ceasing to publish—powerfully affects the global science workforce. Some scientists stay on in academic science and keep publishing, while others stop publishing (Geuna & Shibayama, 2015; Preston, 2004). About one-third disappear from academic publishing within 4 years—which is mostly the time spent in doctoral schools—and about a half within a decade (Kwiek & Szymula, 2023, 2024). Within the first 15 years of academic publishing, those who stay on in science are already distributed among decile-based classes of individual productivity, from the top 10% to the bottom 10%, within their disciplines. Our focus is only on scientists who stay on in science (and keep publishing) for at least 25 years, hence following our definition of late-career scientists in the present research.

What is stunning is the persistence of membership in global top and bottom productivity classes from a life cycle perspective. Later on in their careers, the majority of global top performers (decile 10 in productivity distribution) keep being top performers, and about one-third of global bottom performers (decile 1 in productivity distribution) keep being bottom performers. For them, the probability of staying in global top and bottom productivity classes—horizontal mobility in productivity—over the decades of scientific careers is high; in contrast, the probability of radically vertically changing productivity classes (Jumpers-Up, Droppers-Down) is extremely limited.

The global science system as it emerges from the current research is highly immobile in terms of membership in productivity classes: Jumpers-Up and Droppers-Down are extremely rare (e.g., our microlevel data show that only 0.51% scientists move from early-career bottom class to midcareer top class; and only 0.26% scientists move from early-career top class to midcareer bottom class; there are only 162 and 82 such outliers, respectively, out of 32,063 scientists in all fields combined). The share of Jumpers-Up and Droppers-Down is lower in social science disciplines than in STEMM disciplines.

For instance, among all current top-performing late-career economists and psychologists, there are only four Jumpers-Up: two economists (0.52%) and two psychologists (0.36%) who experienced extreme mobility from decile 1 to decile 10).

The strength of microlevel data lies in the rich insights they offer into the unique trajectories of outlier individuals who have made significant upward leaps in productivity, always of interest in scientometric research and science policy studies. Large-scale computations (e.g., 1.8 billion cited references used to define unique academic discipline for each scientist in our data set) allow us to explore not only these unique cases but also any other individuals or groups in our study. In essence, we can gain a comprehensive understanding—within the constraints of the database and computing methods—of who these outlier scientists are, how they collaborate, publish, and work, and how their contributions are received by the academic community.

Using a longitudinal approach to individual scientists, we know where current global top performers and current global bottom performers come from: As many as 8 in 10 global top performers come from deciles 8–10 (83.66% of midcareer scientists and 83.39% of late-career scientists, a stunningly similar percentage). Analogously, global current bottom performers predominantly come from deciles 1–3 (from 75.31% in the first stage to 68.40% in the second stage), with some cross-disciplinary variation.

Individual research productivity emerges from our regression analyses as highly path-dependent: For all the examined disciplines, STEMM and SOCIAL clusters alike, there is a single most important predictor of becoming a top productive late-career scientist (and a top productive midcareer scientist): being a top productive scientist at an earlier career stage.

The TOP200 predictor (working in the 200 most research-intensive institutions globally) substantially (by 30–50%) increases the odds of success in half of the disciplines studied, with the highest impact in MED, which is the largest discipline, hence pointing to the role of expensive infrastructure in high performance in medical research. Finally, the team size is especially important for membership in the top productive midcareer class in traditionally sole-authored or small-team disciplines, such as the three social science disciplines (BUS, ECON, and PSYCH) and MATH in the STEMM cluster. In these disciplines, a one-unit increase (i.e., one more coauthor in publications in an early-career period) increases the probability of success by 25–40%, testifying to the importance of collaboration in science (Kwiek, 2021; Wagner, 2018).

A general picture of mobility between productivity classes over the course of entire scientific careers based on discipline-aggregated data hides much more nuanced pictures for different disciplines. Some disciplines are much more competitive at the very beginning of their careers, with radical upward mobility being extremely difficult (as in CHEM and COMP, with probabilities of 0.07% and 0.13%, respectively). There are also other disciplines that are much less competitive from the very beginning, in which the presence of Jumpers-Up is much higher (e.g., PHYS, 0.75%, or 10 times more than in CHEM).

A fundamental question emerges from this research: Why do prior productivity class memberships (top, bottom), to a large extent, determine later class memberships (top, bottom)? We can speculate that there are two explanations. First, previous research has shown that the distribution of productivity among scientists has always been highly skewed (Abramo et al., 2017; Albarrán et al., 2011; David, 1994) and that a minority of scientists have always been responsible for the vast majority of publications (Allison, 1980; Ruiz-Castillo & Costas, 2014; Xie, 2014). The old research and science policy theme, which can be summarized as "the majority of scientific work is performed by a relatively small number of scientists" (Crane, 1965, p. 714), has been at the core of theories of individual research productivity. The traditional reward system in science is based on publications. In addition, academic promotions and tenure prospects, salary levels, free time for research, and access to research grants are more or less directly all related to publication productivity (Allison et al., 1982; Stephan, 2012).

Previous success breeds current and future success—here, success being early membership in the tiny class of global top research performers.

Second, higher publishing productivity generally leads to new research funding, as the credibility cycle in academic careers shows (Latour & Woolgar, 1986). In this cycle, research published in prestigious journals (quantity, quality) is converted into recognition; successful grant applications are converted into new equipment, arguments, and articles. The credibility cycle may be more consequential, determining career opportunities in the early-career stages: Once funded based on prestigious articles, scientists' chances of being funded again are higher than those of their less-productive colleagues, at least in more meritocratic national research funding systems with substantial individual-level grant funding. Once funded, and once they have excellent publications, scientists have better odds to be funded again and to be promoted sooner to higher ranks, reflecting the idea that each element of the credibility cycle in academic careers "is but one part of an endless cycle of investment and conversion" (Latour & Woolgar, 1986, p. 200).

In terms of shifting productivity classes from a life cycle perspective, scientists who are less successful early on in their careers (and success here requires a combination of productivity, motivation, determination, aspirations, mentorship, resources, quality of training, innate abilities, and luck) will find it difficult, if not impossible, to prove that they are as good as their more successful, more productive, more determined, more able, better trained, luckier, and possibly better funded colleagues.

Initial publication success is highly correlated with later publication success, which may be explained by two factors: Scientists from the very beginning are different, with some being much more productive, and scientists happen to experience initial publishing success for reasons unrelated to their exceptionality. In both cases, others may view them more positively, leading to more successes in grant acquisition, publishing acceptance, and so forth. Any of the two major explanations separately or together can work for individuals, strengthening the credibility cycle in academic careers.

Unfortunately, in this research, we do not have access to institutional-level and department-level data which are available in survey research designs. We realize that critical to research outcomes are the features of institutional settings in which scientists are actually embedded, as a social-organizational approach suggests (Fox & Nikivincze, 2021). Still, the Scopus data set does not permit us to examine the social dimensions of scientific work. For productivity, both the individual features of scientists and the features of their departments matter. In the majority of individual careers, there is a single institution; sometimes, there are several institutions in entire academic careers.

The features of settings—from the individual level of mentors within departments to the level of whole institutions—may powerfully shape individual career trajectories. Different departments have different "climates": science is a collaborative endeavor and scientists do not operate in a social vacuum (Fox, 2010): In some departmental units research is more welcome than in others, which may shape the mobility/immobility dynamics. Research productivity examined in this paper is strongly linked to the "personality," "character," or "climate" in basic units in which scientists are embedded. As Fox and Mohapatra explain (2007, p. 542), as scientific work takes place in organizations that may either facilitate or inhibit performance, in order to understand scientific productivity, "we need to assess explicitly how social-organizational characteristics of work groups, practices, and climates are associated with performance." This is what research driven by bibliometric data cannot offer.

Social and organizational contexts heavily influence productivity—and yet cannot be grasped in our study. Work climate matters and involves meanings attached to features of organizational life: "work climate can activate interests, convey standards, and stimulate of stifle performance," as Fox and Mohapatra explain (2007, p. 545). Features of settings encompass departmental climates, patterns of interaction among faculty and among faculty and students (such as the practice of frequently speaking with departmental faculty about research—or social integration), the clarity of criteria for tenure and promotion etc. (Fox, 2015; Fox & Nikivincze, 2021).

Bibliometric approaches like ours do not permit analysis of work settings (resources available, work climates in departments) and characteristics such as work practices (Fox & Nikivincze, 2021), as opposed to survey instruments used in academic profession studies. Bibliometric approaches do not permit us to examine and quantify the "personality" of a basic unit in which scientists operate, making some scientists in some units work harder in research and others, in different units, spend less time and energy on research. The consequential links between individuals and their departments cannot be studied in large-scale research like ours. Our approach also does not permit analysis of "who benefits" in collaboration, depending on the rank of scientists, the only proxy of rank being years spent in universities since the first publication indexed in our database.

Our research has reconfirmed the power of a very strong track record as opposed to a very weak track record in science (whenever individual scientists are assessed by research funding panels and promotion committees): For a variety of reasons—which we are not able to fully examine using our data set—the probability of past global top performers becoming global top performers in the future is very high, and their probability of becoming global bottom performers (Droppers-Down) is marginal. At the same time, the chances of global bottom performers reaching the productivity levels achieved by their top-performing colleagues in the very same career stages and within their disciplines (Jumpers-Up) are marginal. Catching up with the global top performers just does not happen, except for a few outliers. (And in some systems, as in Poland, it does not happen at all: The chances we have computed for Poland are 0%; Kwiek & Roszka, 2024b.)

Our findings may have implications for institutional policies, especially regarding hiring and promotions where up or out policies are weak. We show that scientists are heavily locked in to productivity classes early on in their careers. Therefore, hiring and promotion decisions made at the level of departments have long-lasting effects on the productivity of institutions for many years. The chances of institutions that hire and promote predominantly highly productive scientists to have highly productive faculty are very high; at the same time, hiring and promoting bottom productive scientists in fact means tacit institutional agreement to have bottom productive scientists for many years. Some scientists in both STEMM and SOCIAL clusters of disciplines tend to be highly productive for years and decades; and their identification, hiring, and promotion contributes heavily to overall institutional achievements in research. At the same time, while administrative actions of hiring may be shorter-term, the "climates" or "personalities" of departments (units) are costly long-term investments (Fox, 2010, 2015).

Persistent productivity stratification emerges from our individual microlevel analyses as a powerful feature of global science. Using large numbers of observations, our analyses confirm what traditional productivity theories have been claiming for decades, albeit by using small-scale interviews and surveys (Cole & Cole, 1973; Hermanowicz, 2012; Leišytė & Dee, 2012; Merton, 1973): Success breeds success (as in the "cumulative advantage" theory of research productivity), and some scientists will always be globally high performing while others will always be globally low performing (as in the "sacred spark" theory of research productivity).

Our global and longitudinal approach to mobility between research productivity classes at the microlevel of individual scientists uses various proxies and relies on different trade-offs (as discussed in detail in Section 3.6) but hopefully shows new patterns that are, so far, largely underexplored in academic career studies.

*This paper is accompanied by* ESM, *which is available online.*

**ACKNOWLEDGMENTS**

We are grateful to the hosts and audiences of the invited seminars at the University of Oxford (Simon Marginson, CGHE, Center for Global Higher Education, June 2022), Stanford University (John Ioannidis, METRICS, Meta-Research Innovation Center, June 2022 and June 2025), DZHW Berlin (Torger Möller, German Center for Higher Education Research and Science Studies, June 2023), Leiden University (Ludo Waltman, CWTS, Centre for Science and Technology Studies, June 2023), and Sciences Po (Christine Musselin, Sciences Po Paris, October 2023), where Marek Kwiek discussed the strengths and limitations of longitudinal academic career studies using structured Big Data of the Scopus type. We gratefully acknowledge the assistance of the International Center for the Studies of Research (ICSR) Lab, with particular gratitude to Kristy James and Alick Bird from the Lab. We also want to thank Dr Wojciech Roszka from the CPPS Poznan Team for the numerous fruitful discussions and our joint publications about Polish top performers and how to measure their mobility using national data sets.

**AUTHOR CONTRIBUTIONS**

Marek Kwiek: Conceptualization, Data curation, Formal analysis, Investigation, Methodology, Resources, Software, Validation, Writing—original draft, Writing—review & editing. Lukasz Szymula: Conceptualization, Data curation, Formal analysis, Investigation, Methodology, Software, Validation, Visualization, Writing—original draft, Writing—review & editing.

**COMPETING INTERESTS**

The authors have no competing interests.

**FUNDING INFORMATION**

We gratefully acknowledge the support provided by the Ministry of Education and Science through its NDS grant no. NdS/529032/2021/2021.

**DATA AVAILABILITY**

We used microlevel data from Scopus, a proprietary scientometric database. For legal reasons, data from Scopus received through collaboration with the ICSR Lab cannot be made openly available. However, aggregated data can be available from the authors.

**REFERENCES**

Abramo, G., D'Angelo, C. A., & Caprasecca, A. (2009). The contribution of star scientists to overall sex differences in research productivity. *Scientometrics*, *81*, 137–156. https://doi.org/10.1007/s11192-008-2131-7

Abramo, G., D'Angelo, C. A., & Soldatenkova, A. (2017). An investigation on the skewness patterns and fractal nature of research productivity distributions at field and discipline level. *Journal of Informetrics*, *11*(1), 324–335. https://doi.org/10.1016/j.joi.2017.02.001

Aguinis, H., & O'Boyle, E. (2014). Star performers in twenty-first century organizations. *Personnel Psychology*, *67*(2), 313–350. https://doi.org/10.1111/peps.12054

Albarrán, P., Crespo, J. A., Ortuño, I., & Ruiz-Castillo, J. (2011). The skewness of science in 219 sub-fields and a number of

aggregates. *Scientometrics*, *88*(2), 385–397. https://doi.org/10.1007/s11192-011-0407-9
Allison, P. D. (1980). Inequality and scientific productivity. *Social Studies of Science*, *10*, 163–179. https://doi.org/10.1177/030631278001000203
Allison, P. D., Long, J. S., & Krauze, T. K. (1982). Cumulative advantage and inequality in science. *American Sociological Review*, *47*(5), 615–625. https://doi.org/10.2307/2095162
Allison, P. D., & Stewart, J. A. (1974). Productivity differences among scientists: Evidence for accumulative advantage. *American Sociological Review*, *39*(4), 596–606. https://doi.org/10.2307/2094424
Baas, J., Schotten, M., Plume, A., Côté, G., & Karimi, R. (2020). Scopus as a curated, high-quality bibliometric data source for academic research in quantitative science studies. *Quantitative Science Studies*, *1*(1), 377–386. https://doi.org/10.1162/qss_a_00019
Branch, E. H. (Ed.) (2016). *Pathways, potholes, and the persistence of women in science: Reconsidering the pipeline*. Lanham, MD: Lexington Books.
Carrasco, R., & Ruiz-Castillo, J. (2014). The evolution of the scientific productivity of highly productive economists. *Economic Inquiry*, *52*(1), 1–16. https://doi.org/10.1111/ecin.12028
Clauset, A., Larremore, D. B., & Sinatra, R. (2017). Data-driven predictions in the science of science. *Science*, *355*, 477–480. https://doi.org/10.1126/science.aal4217, PubMed: 28154048
Cole, J. R., & Cole, S. (1973). *Social stratification in science*. Chicago, IL: University of Chicago Press.
Costas, R., & Bordons, M. (2007). A classificatory scheme for the analysis of bibliometric profiles at the micro level. In *Proceedings of ISSI 2007: 11th International Conference of the ISSI* (Vols. I and II, pp. 226–230). Madrid: CSIC.
Costas, R., van Leeuwen, T. N., & Bordons, M. (2010). Self-citations at the meso and individual levels: Effects of different calculation methods. *Scientometrics*, *82*, 517–537. https://doi.org/10.1007/s11192-010-0187-7, PubMed: 20234766
Crane, D. (1965). Scientists at major and minor universities: A study of productivity and recognition. *American Sociological Review*, *30*(5), 699–714. https://doi.org/10.2307/2091138
David, P. A. (1994). Positive feedbacks and research productivity in science: Reopening another black box. In O. Granstrand (Ed.), *Economics of technology* (pp. 65–89). Amsterdam: Elsevier.
DiPrete, T. A., & Eirich, G. M. (2006). Cumulative advantage as a mechanism for inequality: A review of theoretical and empirical developments. *Annual Review of Sociology*, *32*, 271–297. https://doi.org/10.1146/annurev.soc.32.061604.123127
Dusdal, J., & Powell, J. J. W. (2021). Benefits, motivations, and challenges of international collaborative research: A sociology of science case study. *Science and Public Policy*, *48*(2), 235–245. https://doi.org/10.1093/scipol/scab010
Fox, M. F. (1983). Publication productivity among scientists: A critical review. *Social Studies of Science*, *13*(2), 285–305. https://doi.org/10.1177/030631283013002005
Fox, M. F. (2010). Women and men faculty in academic science and engineering: Social-organizational indicators and implications. *American Behavioral Scientist*, *53*(7), 997–1012. https://doi.org/10.1177/0002764209356234
Fox, M. F. (2015). Gender and clarity of evaluation among academic scientists in research universities. *Science, Technology, & Human Values*, *40*(4), 487–515. https://doi.org/10.1177/0162243914564074
Fox, M. F., & Mohapatra, S. (2007). Social-organizational characteristics of work and publication productivity among academic scientists in doctoral-granting departments. *Journal of Higher Education*, *78*(5), 542–571. https://doi.org/10.1080/00221546.2007.11772329
Fox, M. F., & Nikivincze, I. (2021). Being highly prolific in academic science: Characteristics of individuals and their departments. *Higher Education*, *81*, 1237–1255. https://doi.org/10.1007/s10734-020-00609-z
Geuna, A., & Shibayama, S. (2015). Moving out of academic research: Why do scientists stop doing research? In A. Geuna (Ed.), *Global mobility of research scientists* (pp. 271–303). Amsterdam: Elsevier. https://doi.org/10.1016/B978-0-12-801396-0.00010-7
Hammarfelt, B. (2017). Recognition and reward in the academy: Valuing publication oeuvres in biomedicine, economics and history. *Aslib Journal of Information Management*, *69*(5), 607–623. https://doi.org/10.1108/AJIM-01-2017-0006
Heckman, J. J., & Moktan, S. (2018). *Publishing and promotion in economics: The tyranny of the Top Five*. NBER Working Paper 25093. https://doi.org/10.3386/w25093
Hermanowicz, J. (2012). The sociology of academic careers: Problems and prospects. In J. C. Smart & M. B. Paulsen (Eds.), *Higher education: Handbook of theory and research* (pp. 207–248). Springer. https://doi.org/10.1007/978-94-007-2950-6_4
Horta, H., & Santos, J. M. (2016). The impact of publishing during PhD studies on career research publication, visibility, and collaborations. *Research in Higher Education*, *57*, 28–50. https://doi.org/10.1007/s11162-015-9380-0
Huang, J., Gates, A. J., Sinatra, R., & Barabási, A.-L. (2020). Historical comparison of gender inequality in scientific careers across countries and disciplines. *Proceedings of the National Academy of Sciences*, *117*(9), 4609–4616. https://doi.org/10.1073/pnas.1914221117, PubMed: 32071248
Ioannidis, J. P. A., Boyack, K. W., & Klavans, R. (2014). Estimates of the continuously publishing core in the scientific workforce. *PLOS ONE*, *9*(7), e101698. https://doi.org/10.1371/journal.pone.0101698, PubMed: 25007173
Jung, J. (2014). Research productivity by career stage among Korean academics. *Tertiary Education and Management*, *20*(2), 85–105. https://doi.org/10.1080/13583883.2014.889206
Kelchtermans, S., & Veugelers, R. (2013). Top research productivity and its persistence: Gender as a double-edged sword. *Review of Economics and Statistics*, *95*(1), 273–285. https://doi.org/10.1162/REST_a_00275
King, M. M., Bergstrom, C. T., Correll, S. J., Jacquet, J., & West, J. D. (2017). Men set their own cites high: Gender and self-citation across fields and over time. *Socius*, *3*. https://doi.org/10.1177/2378023117738903
Kwiek, M. (2016). The European research elite: A cross-national study of highly productive academics across 11 European systems. *Higher Education*, *71*(3), 379–397. https://doi.org/10.1007/s10734-015-9910-x
Kwiek, M. (2018). High research productivity in vertically undifferentiated higher education systems: Who are the top performers? *Scientometrics*, *115*(1), 415–462. https://doi.org/10.1007/s11192-018-2644-7, PubMed: 29527074
Kwiek, M. (2021). What large-scale publication and citation data tell us about international research collaboration in Europe. *Studies in Higher Education*, *46*(12), 2629–2649. https://doi.org/10.1080/03075079.2020.1749254
Kwiek, M. (2023). The globalization of science: The increasing power of individual scientists. In P. Mattei, X. Dumay, E. Mangez, & J. Behrend (Eds.), *The Oxford handbook of education and globalization* (pp. 728–761). Oxford: Oxford University Press. https://doi.org/10.1093/oxfordhb/9780197570685.013.16

Kwiek, M., & Roszka, W. (2024a). Once highly productive, forever highly productive? Full professors' research productivity from a longitudinal perspective. *Higher Education*, *87*, 519–549. https://doi.org/10.1007/s10734-023-01022-y

Kwiek, M., & Roszka, W. (2024b). Are scientists changing their research productivity classes when they move up the academic ladder? *Innovative Higher Education*, *50*, 329–369. https://doi.org/10.1007/s10755-024-09735-3

Kwiek, M., & Szymula, L. (2023). Young male and female scientists: A quantitative exploratory study of the changing demographics of the global scientific workforce. *Quantitative Science Studies*, *4*(4), 902–937. https://doi.org/10.1162/qss_a_00276

Kwiek, M., & Szymula, L. (2024). Quantifying attrition in science: A cohort-based, longitudinal study of scientists in 38 OECD countries. *Higher Education*, *89*, 1465–1493. https://doi.org/10.1007/s10734-024-01284-0

Kwiek, M., & Szymula, L. (2025). Leaving science—Attrition of biologists in 38 OECD countries. *FEBS Letters*, *599*(6), 799–812. https://doi.org/10.1002/1873-3468.70028, PubMed: 40055998

Larivière, V., Ni, C., Gingras, Y., Cronin, B., & Sugimoto, C. R. (2013). Bibliometrics: Global gender disparities in science. *Nature*, *504*(7479), 211–213. https://doi.org/10.1038/504211a, PubMed: 24350369

Latour, B. & Woolgar, S. (1986) *Laboratory life: The construction of scientific facts*. Princeton, NJ: Princeton University Press. https://doi.org/10.1515/9781400820412

Leišytė, L., & Dee, J. R. (2012). Understanding academic work in a changing institutional environment. In J. C. Smart & M. B. Paulsen (Eds.), *Higher education: Handbook of theory and research* (pp. 123–206). Springer. https://doi.org/10.1007/978-94-007-2950-6_3

Li, W., Aste, T., Caccioli, F., & Livan, G. (2019). Early coauthorship with top scientists predicts success in academic careers. *Nature Communications*, *10*, 5170. https://doi.org/10.1038/s41467-019-13130-4, PubMed: 31729362

Lindahl, J. (2018). Predicting research excellence at the individual level: The importance of publication rate, top journal publications, and top 10% publications in the case of early career mathematicians. *Journal of Informetrics*, *12*(2), 518–533. https://doi.org/10.1016/j.joi.2018.04.002

Liu, L., Jones, B. F., Uzzi, B., & Wang, D. (2023). Data, measurement and empirical methods in the science of science. *Nature Human Behaviour*, *7*(7), 1046–1058. https://doi.org/10.1038/s41562-023-01562-4, PubMed: 37264084

Lutter, M., & Schröder, M. (2016). Who becomes a tenured professor, and why? Panel data evidence from German sociology, 1980–2013. *Research Policy*, *45*(5), 999–1013. https://doi.org/10.1016/j.respol.2016.01.019

Ma, Y., Mukherjee, S., & Uzzi, B. (2020). Mentorship and protégé success in STEM fields. *Proceedings of the National Academy of Sciences*, *117*(25), 14077–14083. https://doi.org/10.1073/pnas.1915516117, PubMed: 32522881

Marginson, S. (2022). What drives global science? The four competing narratives. *Studies in Higher Education*, *47*(8), 1566–1584. https://doi.org/10.1080/03075079.2021.1942822

Menard, S. (2002). *Longitudinal research*. Thousand Oaks, CA: Sage. https://doi.org/10.4135/9781412984867

Merton, R. K. (1973). *The sociology of science: Theoretical and empirical investigations*. Chicago, IL: University of Chicago Press.

Ni, C., Smith, E., Yuan, H., Larivière, V., & Sugimoto, C. R. (2021). The gendered nature of authorship. *Science Advances*, *7*(36), eabe4639. https://doi.org/10.1126/sciadv.abe4639, PubMed: 34516891

Nielsen, M. W., & Andersen, J. P. (2021). Global citation inequality is on the rise. *Proceedings of the National Academy of Sciences*, *118*(7), e2012208118. https://doi.org/10.1073/pnas.2012208118, PubMed: 33558230

Nygaard, L. P., Aksnes, D. W., & Piro, F. N. (2022). Identifying gender disparities in research performance: The importance of comparing apples with apples. *Higher Education*, *84*, 1127–1142. https://doi.org/10.1007/s10734-022-00820-0

Preston, A. E. (2004). *Leaving science: Occupational exit from scientific careers*. New York, NY: Russell Sage Foundation.

Priem, J., Piwowar, H., & Orr, R. (2022). OpenAlex: A fully-open index of scholarly works, authors, venues, institutions, and concepts. *arXiv*. https://doi.org/10.48550/arXiv.2205.01833

Rørstad, K., & Aksnes, D. W. (2015). Publication rate expressed by age, gender and academic position—A large-scale analysis of Norwegian academic staff. *Journal of Informetrics*, *9*(2), 317–333. https://doi.org/10.1016/j.joi.2015.02.003

Ross, M. B., Glennon, B. M., Murciano-Goroff, R., Berkes, E. G., Weinberg, B. A., & Lane, J. I. (2022). Women are credited less in science than men. *Nature*, *608*(7921), 135–145. https://doi.org/10.1038/s41586-022-04966-w, PubMed: 35732238

Rowland, D. T. (2014). *Demographic methods and concepts*. Oxford: Oxford University Press.

Ruiz-Castillo, J., & Costas, R. (2014). The skewness of scientific productivity. *Journal of Informetrics*, *8*(4), 917–934. https://doi.org/10.1016/j.joi.2014.09.006

Ruspini, E. (1999). Longitudinal research and the analysis of social change. *Quality and Quantity*, *33*(3), 219–227. https://doi.org/10.1023/A:1004692619235

Salganik, M. J. (2018). *Bit by bit: Social research in the digital age*. Princeton, NJ: Princeton University Press.

Savage, W. E., & Olejniczak, A. J. (2021). Do senior faculty members produce fewer research publications than their younger colleagues? Evidence from Ph.D. granting institutions in the United States. *Scientometrics*, *126*, 4659–4686. https://doi.org/10.1007/s11192-021-03957-4

Shibayama, S., & Baba, Y. (2015). Impact-oriented science policies and scientific publication practices: The case of life sciences in Japan. *Research Policy*, *44*(4), 936–950. https://doi.org/10.1016/j.respol.2015.01.012

Shin, J. C., & Cummings, W. K. (2010). Multilevel analysis of academic publishing across disciplines: Research preference, collaboration, and time on research. *Scientometrics*, *85*, 581–594. https://doi.org/10.1007/s11192-010-0236-2

Spoon, K., LaBerge, N., Wapman, K. H., Zhang, S., Morgan, A. C., ... Clauset, A. (2023). Gender and retention patterns among U.S. faculty. *Science Advances*, *9*(42), eadi2205. https://doi.org/10.1126/sciadv.adi2205, PubMed: 37862417

Stephan, P. (2012). *How economics shapes science*. Cambridge, MA: Harvard University Press. https://doi.org/10.4159/harvard.9780674062757

Stephan, P. E., & Levin, S. G. (1992). *Striking the mother lode in science: The importance of age, place, and time*. Oxford: Oxford University Press.

Sugimoto, C., & Larivière, V. (2018). *Measuring research: What everyone needs to know*. Oxford: Oxford University Press. https://doi.org/10.1093/wentk/9780190640118.001.0001

Sugimoto, C., & Larivière, V. (2023). *Equity for women in science: Dismantling systemic barriers to advancement*. Cambridge, MA: Harvard University Press. https://doi.org/10.4159/9780674292918

Thelwall, M., & Sud, P. (2022). Scopus 1900–2020: Growth in articles, abstracts, countries, fields, and journals. *Quantitative Science Studies*, *3*(1), 37–50. https://doi.org/10.1162/qss_a_00177

Turner, L., & Mairesse, J. (2005). *Individual productivity differences in public research: How important are non-individual determinants? An econometric study of French physicists' publications and citations (1986–1997)*. Paris: CNRS.

Wagner, C. S. (2018). *The collaborative era in science: Governing the network*. London: Palgrave Macmillan. https://doi.org/10.1007/978-3-319-94986-4

Wang, D., & Barabási, A.-L. (2021). *The science of science*. Cambridge: Cambridge University Press. https://doi.org/10.1017/9781108610834

Wang, Y., Jones, B. F., & Wang, D. (2019). Early career setback and future career impact. *Nature Communications*, *10*, 4331. https://doi.org/10.1038/s41467-019-12189-3, PubMed: 31575871

Way, S. F., Morgan, A. C., Clauset, A., & Larremore, D. B. (2017). The misleading narrative of the canonical faculty productivity trajectory. *Proceedings of the National Academy of Sciences*, *114*(44), E9216–E9223. https://doi.org/10.1073/pnas.1702121114, PubMed: 29042510

Weingart, P. (2004). Impact of bibliometrics upon the science system: Inadvertent consequences? In H. F. Moed, W. Glänzel, & U. Schmoch (Eds.), *Handbook on quantitative science and technology research* (pp. 117–131). Amsterdam: Kluwer Academic Publishers.

West, J. D., Jacquet, J., King, M. M., Correll, S. J., & Bergstrom, C. T. (2013). The role of gender in scholarly authorship. *PLOS ONE*, *8*(7), e66212. https://doi.org/10.1371/journal.pone.0066212, PubMed: 23894278

Xie, Y. (2014). "Undemocracy": Inequalities in science. *Science*, *344*(6186), 809–810. https://doi.org/10.1126/science.1252743, PubMed: 24855244

Zeng, A., Shen, Z., Zhou, J., Wu, J., Fan, Y., … Stanley, H. E. (2017). The science of science: From the perspective of complex systems. *Physics Reports*, *714–715*, 1–73. https://doi.org/10.1016/j.physrep.2017.10.001

Zhang, S., Wapman, K. H., Larremore, D. B., & Clauset, A. (2022). Labor advantages drive the greater productivity of faculty at elite universities. *Science Advances*, *8*(46), eabq7056. https://doi.org/10.1126/sciadv.abq7056, PubMed: 36399560